\PassOptionsToPackage{hyphens}{url}

\documentclass[trackchanges]{aastex701}
\usepackage[utf8]{inputenc}
\usepackage[T1]{fontenc}
\usepackage{textcomp}
\usepackage{enumitem}

\begin{document}

\title{Probing the CME Core--Prominence Relation Using Inner Coronal Observations}

\author[orcid=0009-0005-6070-9139,gname=Sunit Sundar, sname=Pradhan]{Sunit Sundar Pradhan}
\affiliation{Indian Institute of Astrophysics, Koramangala, Bangalore 560034, India}
\affiliation{Pondicherry University, R.V. Nagar, Kalapet 605014, Puducherry, India}
\email{sunit.pradhan@iiap.res.in}  

\author[orcid=0000-0003-0585-7030,gname=Jayant, sname=Joshi]{Jayant Joshi} 
\affiliation{Indian Institute of Astrophysics, Koramangala, Bangalore 560034, India}
\affiliation{Pondicherry University, R.V. Nagar, Kalapet 605014, Puducherry, India}
\email{jayant.joshi@iiap.res.in}

\author[orcid=0000-0002-9667-6392,gname=Tanmoy,sname=Samanta]{Tanmoy Samanta}
\affiliation{Indian Institute of Astrophysics, Koramangala, Bangalore 560034, India}
\affiliation{Pondicherry University, R.V. Nagar, Kalapet 605014, Puducherry, India}
\email{tanmoy.samanta@iiap.res.in}

\begin{abstract}

Coronal mass ejections (CMEs) often exhibit a three-part structure consisting of a bright inner core, an outer leading edge, and an intervening dark cavity. While the core has traditionally been attributed to prominence material, an alternative interpretation suggests it may arise from the projection effects of a twisted flux rope.
We focused on limb CME events to reassess the connection between CME cores and their associated prominences in the inner corona. The CME cores were analyzed using white-light observations from the Mauna Loa Solar Observatory (MLSO) K-Coronagraph (K-Cor), while the corresponding prominence eruptions were examined using H$\alpha$ data from the Global Oscillation Network Group (GONG) and 304 \AA{} images from the Atmospheric Imaging Assembly (AIA).
Our results show a strong spatial correspondence between H$\alpha$ prominences and CME cores in white light, with an average image correlation of $\sim$0.7, while correlations between white light and AIA 304 \AA{} are comparatively weaker ($\sim$0.5). Several events could be continuously traced into the Large Angle and Spectrometric Coronagraph Experiment (LASCO/C2) field of view, confirming the persistence of prominence material into the outer corona.
We find back-extrapolating LASCO/C2 CME cores under constant-velocity, linear-trajectory assumptions can introduce large errors—up to $\sim$40\textdegree\ in inferred position angle and $\sim$140 minutes in eruption time relative to their true values—underscoring the importance of inner-coronal observations for accurately constraining CME dynamics.
Overall, our findings suggest that in prominence-associated CMEs, the bright cores are predominantly composed of prominence material.

\end{abstract}


\section{Introduction} 
\label{sec:intro}

Coronal mass ejections (CMEs) are powerful eruptions of plasma and magnetic field from the solar surface into interplanetary space. 
CMEs are typically observed using white-light coronagraphs, which employ one or more occulting disks to block the intense light from the solar disk while allowing the much fainter light from the solar corona to be captured \citep{lyot1930}. 
Since the advent of coronagraphic instruments, our understanding of CMEs and their impact on Earth has undergone a significant leap \citep{Gopalswamy_2009, chen_2011, Webb2012, Temmer_2021}. 
In coronagraph images, CMEs can appear in various shapes.
One common type is called the “three-part” structure, which consists of a bright inner core, a dark cavity in the middle, and an outer leading edge \citep{Illing_1986}. 
The three-part structure is often considered the standard structure of CMEs \citep{chen_2011}. 
However, only about one-third of CMEs show the classical three-part structure \citep{Munro_1979, Webb1987, Vourildas_2013}. 
But interestingly, recent work by \cite{Song_2023_1} presents a different picture: almost all CMEs wider than 60\textdegree\, show a three-part structure in the inner coronal observations from the Mauna Loa Solar Observatory (MLSO) K-Coronagraph \citep[K-Cor,][]{KCORpaepr2016}, i.e., below 3 R$_\odot$. 
Their results underscore the importance of inner coronal observations in fully understanding the three-part structure of CMEs and their evolution.

Numerous observational and modeling efforts in the past have contributed to understanding the nature of the three-part structure, broadly converging into a unified view which is now referred to as the ‘traditional view’ \citep{chen_2011, Webb2012}. 
According to this view, the core of the three-part CME structure can be interpreted as the erupting prominence \citep{House_1981, Illing_1985, Illing_1986}, while the surrounding dark cavity is interpreted as a magnetic flux rope (MFR) \citep{Forbes_2000, Gibson_2006, Fuller_2008, howard_deforest_2012}. 
As a prominence, enclosed within an MFR, rises through the corona, it pushes aside the ambient plasma, forming a bright leading edge of a CME \citep{Forbes_2000, Schwenn_2006, Vourildas_2013}.
However, recently, this perspective has been questioned \citep{Howard_2017, Song_2017,song_2022}, particularly regarding the interpretation of the CME core and its association with prominences. Below, we briefly review the studies that established the traditional view of the core as prominence material, and then outline the recent challenges it has faced.

Following the discovery of CMEs \citep{Hansen_1971}, numerous studies explored their association with solar disk phenomena such as prominences and filaments. Using Skylab \citep{Skylab_1974} data, \cite{Poland_1976} showed that CME core emission was dominated by H$\alpha$, indicating the presence of prominence material. Observations from the Solar Maximum Mission \citep[SMM,][]{SMM1980}, which provided overlapping coronagraph and H$\alpha$ coverage, further revealed that CME cores frequently contained eruptive prominence signatures \citep{House_1981}. A detailed analysis by \cite{Illing_1986}, after separating H$\alpha$ emission and Thomson-scattered light, confirmed the presence of prominence material in CME cores, thereby strengthening the prominence–CME connection.

Additionally, many coronagraphic observations reveal dense, filamentary structures within CME cores \citep{Song_2019_1}, often interpreted as embedded prominence material. Several studies \citep{Schmahl_1977, Burlaga_1998, Gopalswwamy_1998, Skoug_1999, Maricic2004, Mierla_2011, susino_2018} have reported cases where prominence material was directly tracked into CME cores. Together, these findings established the long-standing view that prominences represent CME cores \citep{Webb2012}.

It is important to note, however, that most of these early works were case studies rather than broad statistical analyses. Subsequent studies based on statistical analyses have examined the occurrence rates of eruptive and non-eruptive prominences that lead to CMEs \citep{Wang_1998, Gilbert_2000, Hori_2002, Gopalswamy_2003, Choudhary_2003, Flippov_2008, Maricic2009, Yan_2011, Seki_2021, Schmieder_2013}, as well as the overall prominence–CME association rate \citep{Munro_1979, Webb1987, Subramanian_2001, wang_2011, Satabdwa_2023_1}. However, none of these studies specifically focused on CMEs exhibiting the three-part structure. One notable study is the 16-year statistical analysis of Large Angle and Spectrometric Coronagraph Experiment \citep[LASCO,][]{LASCO_1995} observations by \cite{Vourildas_2013}, which examined the morphological types of CMEs. The study showed that CME cores are generally associated with prominence material. Nonetheless, most such studies are limited by the lack of simultaneous structural overlap between white-light and H$\alpha$/EUV observations, which is essential for reliably linking prominences with CME cores.

One of the first direct challenges to the traditional view came from \cite{Howard_2017}. They compared LASCO/C2 three-part CMEs with nearby prominence observations in the Atmospheric Imaging Assembly \citep[AIA,][]{AIA_2012}. Their study showed that the cores of most three-part CMEs had no clear link to prominences. They also reported unexpectedly low detection rates of prominence material at greater heliocentric distances, both in in-situ measurements \citep{Lepri_2010} and remote-sensing data \citep{Wood_2016}. Moreover, their observations from the COR1 coronagraph \citep{cor1_2003} on the Solar Terrestrial Relations Observatory (STEREO) showed that some CME cores had high polarized brightness–to–total brightness (pB/B) ratios.
The observed values were greater than those anticipated for a core made of cool, dense prominence plasma.
To account for these discrepancies, \cite{Howard_2017} proposed several scenarios. One possibility is that CMEs are generally not associated with prominences. Another is that some events involve failed prominence eruptions, where the material falls back to the solar surface \citep{Gilbert_2007, Filippov_2020}. A further explanation is that prominence plasma may undergo rapid ionization at higher altitudes \citep{Illing&athay_1986, Howard_2015}. \cite{Howard_2017} also proposes an alternative interpretation of the classic three‑part CME structure that does not require a prominence: the entire CME is a flux rope, and the core is a twisted section of that flux rope whose brightness is enhanced by line‑of‑sight (LOS)–integrated emission. However, this study also lacked overlapping FOV observations between white-light and EUV observations.

\citet{Song_2023_2} proposed another alternative perspective, suggesting that in prominence-unrelated CMEs the plasma within the magnetic flux rope appears as the bright core, the plasma pileup along overlying coronal loops forms the bright leading edge, and the intervening low-density region—distinct from the flux rope in the inner corona—corresponds to the dark cavity. In such cases, flux rope plasma is often observed in hot EUV channels (e.g., AIA 94~\AA{} and AIA 131~\AA{}) as a hot-channel structure \citep{Zhang_2012,Song_2014}. 
For prominence-associated CMEs, both prominence material and the magnetic flux rope can contribute to the observed core structure. Coronagraphic and EUV observations show that the sharp, brighter component of the core is associated with erupting prominence material, while a surrounding, more diffuse component corresponds to the flux rope \citep{ Song_2019_1, song_2022}. It was further suggested that the three-part CME structure gradually evolves from the inner to the outer corona \citep{Song_2017, Song_2023_1, Song_2023_2, Song_2025}. In this framework, the flux rope initially confined to the CME core in the inner corona expands outward and forms the dark cavity at larger heights, reconciling inner-coronal interpretations with the traditional three-part picture observed in the outer corona. This view also explains why some prominence-associated CMEs lose their three-part morphology in the outer corona when the prominence material drains back during the eruption \citep{Song_2023_1}.


Considering all these different interpretations, we simply revisit a fundamental question: When a CME is associated with a prominence, does the core consist of that prominence material? In this study, we aim to answer this question through direct, overlapping observations using white light, H$\alpha$, and 304~\AA{}. In many cases, we use LASCO/C2 observations to extend our analysis into the outer corona (beyond 3 R$_\odot$) and follow the evolution of CMEs outside the inner coronal FOV.

\begin{figure*}[ht!]
\centering
\includegraphics[width=0.67\textwidth]{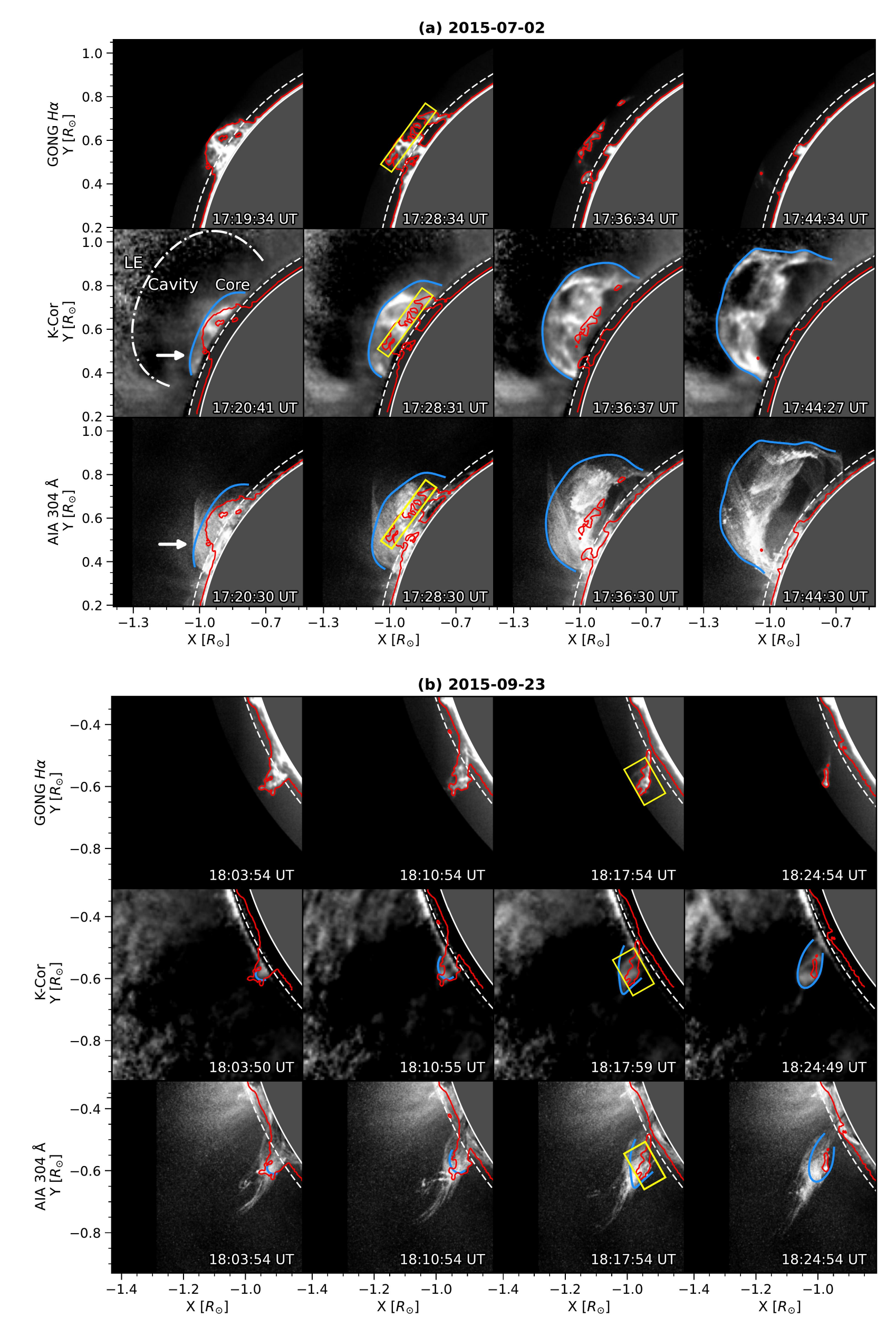}  
\caption{Comparison of the structural evolution of CME cores in K-Cor white-light images with their associated prominences in GONG H$\alpha$ and AIA 304~\AA{} is presented for Category-A CMEs, which preserve their three-part structure into the LASCO/C2 FOV. Panel (a) corresponds to the event on 2015 July 02, while panel (b) shows the event on 2015 September 23. For each event, images from all instruments at four selected time steps are presented to illustrate the evolution of the eruption.
Red contours, outlining features from H$\alpha$ observations, are projected onto both K-Cor and AIA 304~\AA{} images to assess morphological correspondence. Blue contours, extracted from K-Cor structures, are overlaid on the AIA 304~\AA{} images to facilitate further structural comparison. Yellow boxes in the images indicate ROIs selected for detailed analysis and are shown only for timestamps in which the prominence is the most prominent in H$\alpha$. 
A gray area with a solid white outline denotes the solar disk and the limb, respectively.
A dashed white arc in all images marks 1.05 $R_\odot$, the inner edge of the K-Cor FOV. In the first K-Cor frame of panel (a), a white dot-dashed curve indicates the inner boundary of the CME leading edge. The arrows mark regions where a feature seen in white-light and 304~\AA{} is absent in H$\alpha$, which is discussed in detail in Section~\ref{sec:res1}. An animation of this figure is available \href{https://drive.google.com/file/d/1b9A55YyB5nHH3rBnGO941-MuenO235IP/view?usp=sharing}{online}.
\label{2july15sept}}
\end{figure*}

\begin{figure*}[ht!]
\centering
\includegraphics[width=0.67\textwidth]{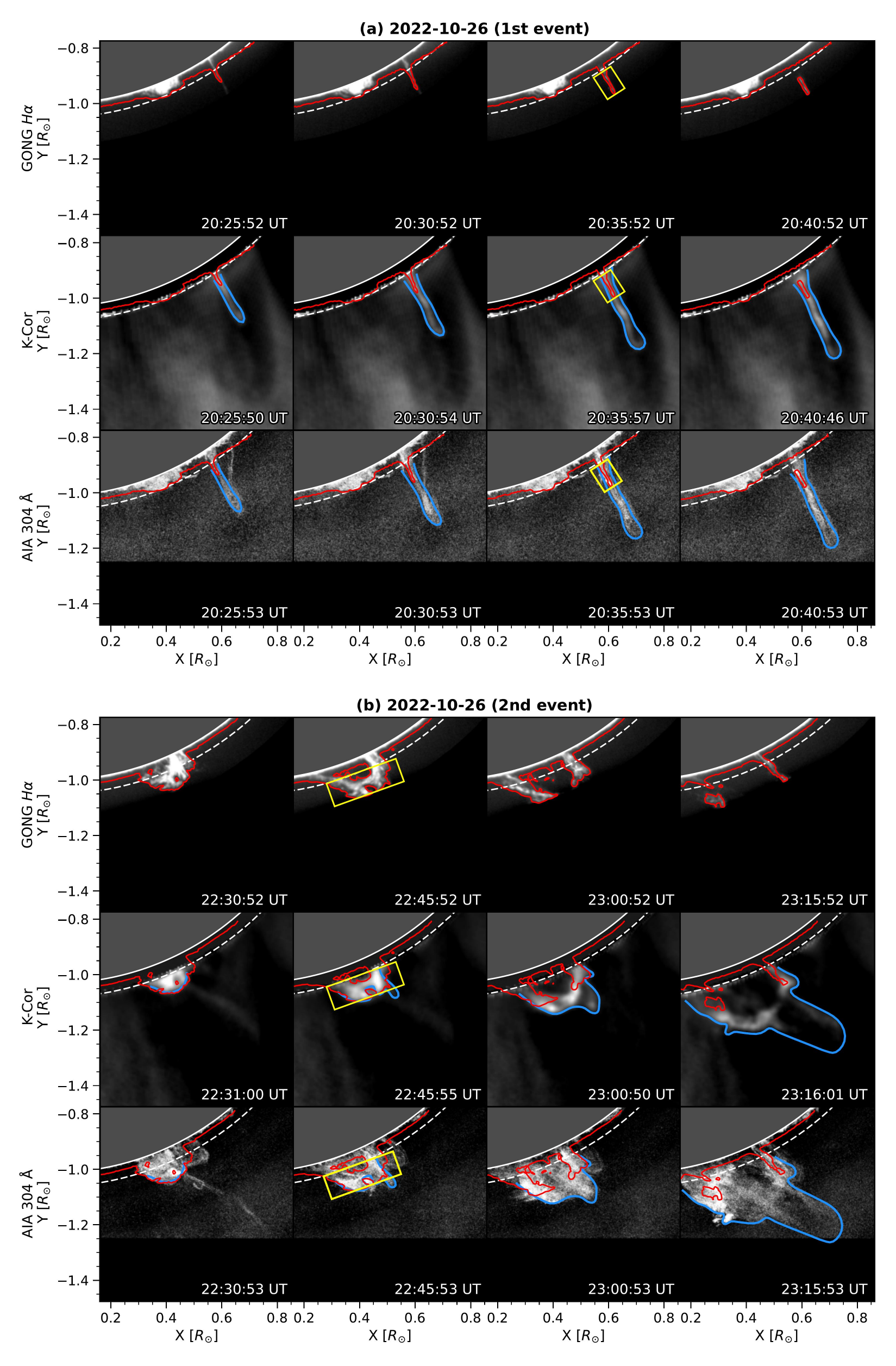}  
\caption{Same as Figure~\ref{2july15sept}, with panels (a) and (b) showing structural analysis of two consecutive Category-A CME events on 2022 October 26.
An animation of this figure is available \href{https://drive.google.com/file/d/1jlqaYhYeLJS4_LzJnHbHs66Zd2eaVkQi/view?usp=sharing}{online}. 
\label{26oct}}
\end{figure*}

\begin{figure*}[ht!]
\centering
\includegraphics[width=1\textwidth]{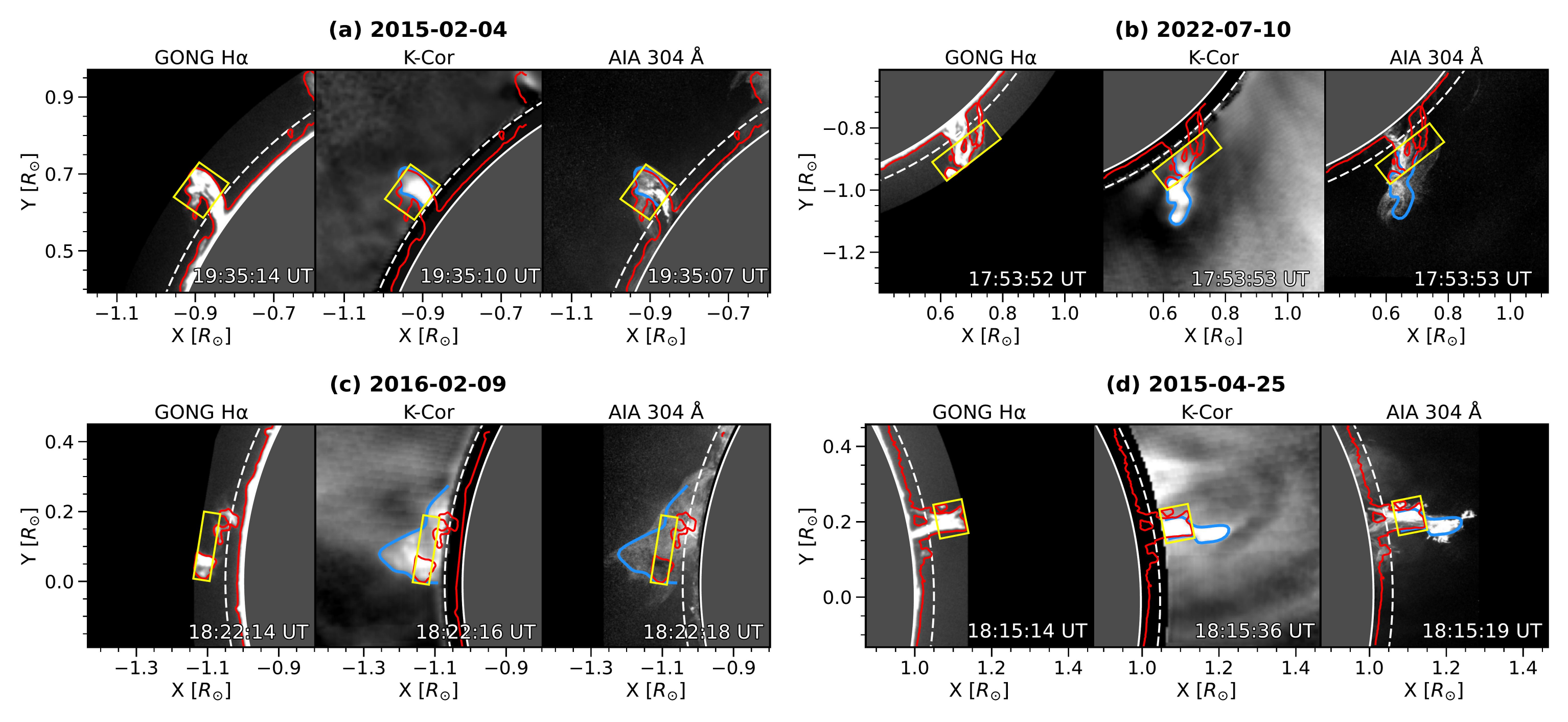}  
\caption{Structural comparison of Category-B CMEs that fail to maintain their three-part structure into the LASCO/C2 FOV is shown at a single representative time step for four distinct events, dated 2015 February 4, 2022 July 10, 2016 February 9, and 2015 April 25, in panels (a)–(d), respectively. The representation of the observational images follows the same format as in Figure~\ref{2july15sept}.
An animation of this figure is available \href{https://drive.google.com/file/d/1chP144_UOntUvH3w-zt0g2CbDGqRxQ_H/view?usp=sharing}{online}.
\label{4x4prominence}}
\end{figure*}

\section{Observations and Data} \label{sec:data}
\subsection{Data Products and Processing} \label{sec:data1}

To facilitate a direct comparison between CME cores and prominences in the lower corona, we utilize observations from the MLSO K-Cor coronagraph.
K-Cor offers the field of view closest to the solar disk among currently available coronagraphs, covering a radial range from approximately 1.05 to 3 R$_\odot$.
It is important to note that the inner boundary can vary slightly between events due to pointing uncertainties inherent to coronagraphic instrumentation.
The instrument operates within a passband of 7200–7500~\AA{}, capturing the polarized brightness (pB) resulting from Thomson scattering of photospheric light by coronal electrons \citep{Van_1950, Billing_1966}.
K-Cor has a plate scale of 5\farcs643 per pixel and provides high-cadence observations with a 15-second temporal resolution.

To observe prominences, we utilized data from AIA 304~\AA{}, Global Oscillation Network Group \citep[GONG,][]{GONG_2011} H$\alpha$, and the 304~\AA{} channel of the Sun–Earth Connection Coronal and Heliospheric Investigation (SECCHI)/Extreme Ultraviolet Imager \citep[EUVI,][]{stereo_euvi_2004}.
AIA is an Extreme Ultraviolet (EUV) imaging instrument aboard the Solar Dynamics Observatory (SDO).
It captures high-resolution images with a plate scale of 0\farcs6 per pixel and a temporal cadence of 12 seconds.
Its FOV extends up to 1.3 R$_\odot$ along the sides.
GONG H-Alpha is a network of six ground-based telescopes strategically positioned around the globe to provide nearly continuous solar observations.
GONG H-Alpha operates with a narrow passband of 0.4–0.5~\AA{} centered at 6562.8~\AA{}, achieving a plate scale of approximately 1\arcsec{} per pixel and a cadence of one minute.
Its FOV covers up to 1.13 R$_\odot$.
EUVI, part of the SECCHI suite onboard the Solar Terrestrial Relations Observatory (STEREO), provides additional EUV observations with a plate scale of approximately 1\farcs59 per pixel and a cadence of 2.5 minutes.
To track CME evolution beyond the inner corona, we also use LASCO/C2 observations from Solar and Heliospheric Observatory \citep[SOHO,][]{SOHO_1995}, which cover a radial range of 2–6 R$_\odot$ in white light, with a plate scale of 11\farcs2 per pixel and a cadence of about 12 minutes.

CME intensity naturally decreases with heliocentric distance; the signal-to-noise ratio (SNR) becomes a limiting factor, particularly in ground-based observations such as those from K-Cor, which are further affected by atmospheric interference.
Although K-Cor has a nominal FOV extending to 3 R$_\odot$, features beyond about 2–2.3 R$_\odot$ tend to be faint and noisy, reducing their utility for detailed analysis.
In this outer range, bright structures may be visible in motion, but their fine-scale features are difficult to discern.
For this study, we used K-Cor data processed with the Normalized Radial Gradient Filter \citep[NRGF,][]{Morgan_2006}, available online\footnote{\href{https://www2.hao.ucar.edu/mlso}{https://www2.hao.ucar.edu/mlso}}.
In some cases, where NRGF-processed images did not provide the desired contrast and visibility of dynamic structures, we performed base differencing by subtracting a background image created using images obtained prior to eruptions.

For H$\alpha$ and 304~\AA{}, we have analyzed only the off-limb part of the images.
H$\alpha$ images were primarily obtained from the GONG station at MLSO, Hawaii. 
When data from the MLSO site were unavailable or of poor quality, observations from other stations—namely, Big Bear Solar Observatory (California) and Cerro Tololo Inter-American Observatory (Chile)—were used instead.
For AIA 304~\AA{} data, a radial filter was applied to enhance fainter structures away from the solar limb. 
No additional processing was applied to the H$\alpha$ images.

Structural comparison posed a challenge due to the limited overlap in the fields of view between instruments. The region of overlap, restricted to approximately 1.05–1.13 R$_\odot$, spans the inner edge of K-Cor and the outer edge of GONG H$\alpha$.
Observations from different instruments were co-aligned by centering on the solar disk and standardizing radial distances to solar radii (R$_\odot$), allowing consistent spatial comparison.
Even after the above-mentioned alignment, we found some residual spatial offset in some cases, with a maximum offset of 40\arcsec{} between instruments, likely due to pointing inaccuracies.
These offsets were manually corrected by comparing identifiable features off the disk for the different times across all datasets.
While this manual alignment introduced potential bias, simultaneous analysis of structures in three complementary instruments improved overall confidence in the results.

To trace the cores of CMEs beyond K-Cor FOV, we supplemented K-Cor data with LASCO/C2 observations from SOHO, which cover a range of ~2–6 R$_\odot$ in white light.
Typically, LASCO/C2 data are reliable only beyond 2.1–2.3 R$_\odot$, as the region closer to the occulter suffers from significant scattered light from the solar disk.
We processed level-0.5 LASCO/C2 data to level-1 using the \texttt{reduce\_level\_one.pro} routine available in SolarSoft.
To enhance faint structures, we applied the Simple Radial Gradient Filter \citep[SiRGraF,][]{Ritesh_2022}, which has been shown to outperform NRGF for outer coronal features.
SiRGraF method involves generating a background image from long-duration observations, deriving a uniform 2D radial profile, and then enhancing the image by subtracting the background and dividing by the profile—highlighting subtle, dynamic features in the outer corona.

\subsection{Event Selection} \label{sec:data2}

    The event selection for this study was inspired by the dataset used in \citet{Song_2023_1}, which focused on CMEs exhibiting a three-part structure within the inner corona as observed by K-Cor. Also, some additional CME events were incorporated based on the following selection criteria:
    \begin{enumerate}[label=\roman*.] 
    \item The CME must display a clear three-part structure in K-Cor observations.
    \item The event should be observed near the solar limb.
    \item The CME must be associated with a prominence. \end{enumerate}
    
Limb events are defined as follows: for Earth-facing events, the source region must be located at least 30\textdegree\, in longitude from the solar limb, and for backside events, the associated prominence structure must be visible \citep{Song_2023_1}. 
They are preferred for their clearer three-part structure and reduced projection effects.

To identify potential events, we referred to the K-Cor CME catalog, which lists detections from 2013 to 2022 based on an automated CME detection algorithm \citep{kcor_2017}.
A total of 65 events were selected based on the criteria outlined above. 
The association with prominences was verified using AIA EUV data, with additional data from the EUVI instrument on SECCHI/STEREO. Each event was visually inspected to determine whether a distinct prominence structure was visible in H$\alpha$ observations, allowing for a meaningful comparison with the core features observed in K-Cor data.
    
Out of the 65 events, 5 did not exhibit any identifiable prominence structure in H$\alpha$. 
In 13 events, a well-defined prominence was initially visible, but it faded quickly and was no longer discernible within the K-Cor FOV during the evolution. 
Additionally, 8 more events were discarded, which did not suffer from the above limitations but were excluded due to other data-related issues.
For example, the unavailability of K-Cor observations because the event took place before K-Cor observations began or because observations were excessively bright near the K-Cor occulter, which obscured the core structure. 
Excluding these 26 events, the remaining 39 showed a clearly defined prominence structure in H$\alpha$ and were included in the detailed analysis.

Among these 39 events, 16 retained a recognizable three-part structure during their evolution into the LASCO/C2 FOV, while the remaining 23 did not. 
Out of the 16 events, one (2014 February 11) was excluded, as a nearby hot channel eruption made it difficult to determine whether the CME core arose from the prominence, the hot channel, or both, leaving 15 events. 
For this study, CMEs that maintain their three-part structure into the LASCO/C2 FOV are classified as Category A, while those that do not are referred to as Category B. 
Table~\ref{tab:taba} lists all 15 Category A CMEs, and Table~\ref{tab:tabb} lists all 23 Category B events.
Additionally, it is worth noting that among the 26 events excluded from detailed analysis, 3 were observed to exhibit a three-part structure in the higher corona as seen in LASCO/C2 observations.

\begin{figure*}[ht!]
\centering
\includegraphics[width=1\textwidth]{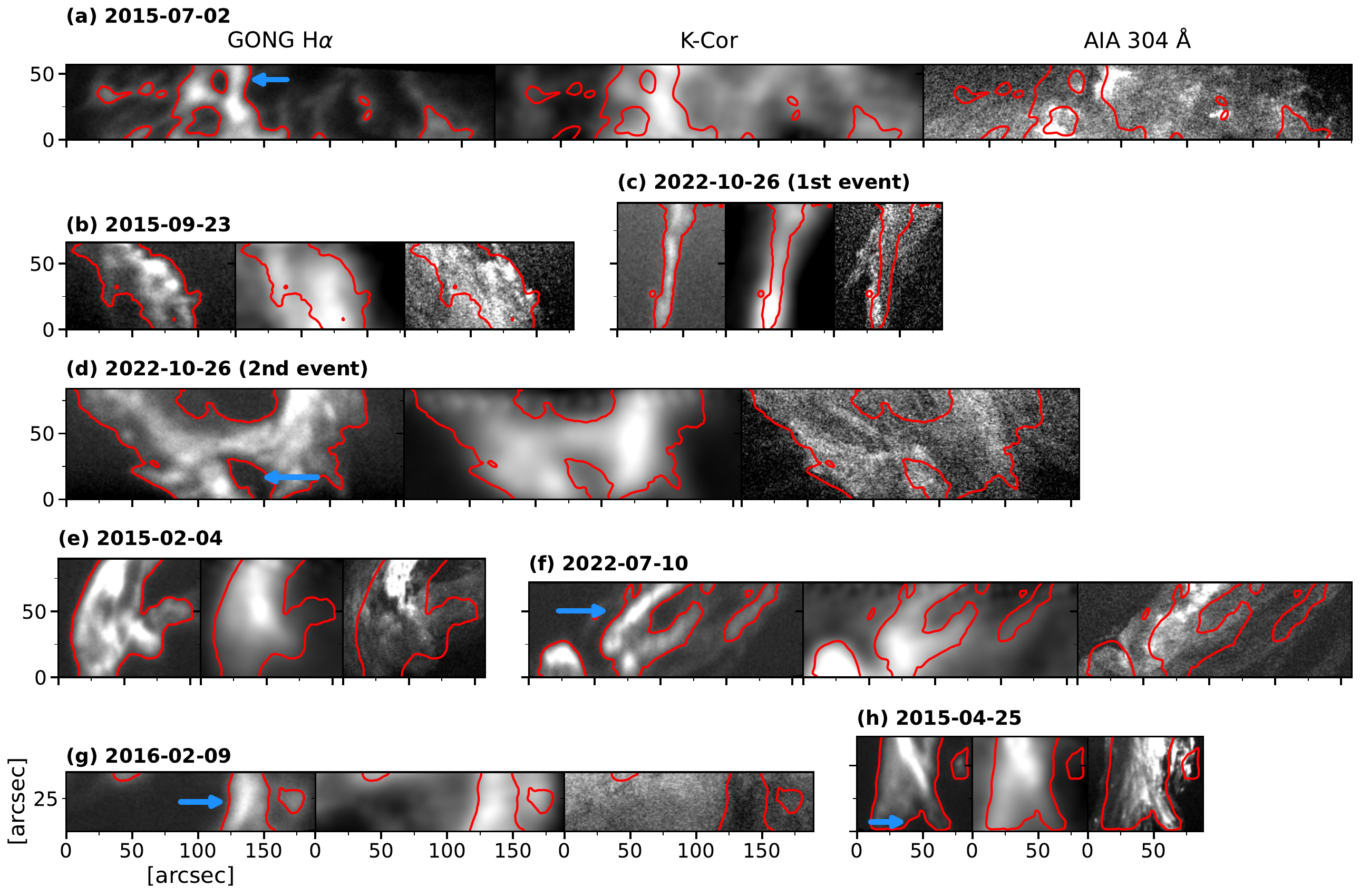}  
\caption{Cutout images of the yellow-box ROIs from GONG H$\alpha$, K-Cor, and AIA 304~\AA{} observations, as shown in Figures~\ref{2july15sept}, \ref{26oct}, and \ref{4x4prominence}. Panels (a)–(d) represent four Category-A CME events, and panels (e)–(h) represent four Category-B CME events. These zoomed-in views highlight the fine-scale structural correspondence between prominence and CME core features across different wavelengths. In each panel, the first image corresponds to GONG H$\alpha$, the second to the K-Cor and the third to the AIA 304 \AA{}. Red contours, extracted from H$\alpha$ images, are overlaid on the 304~\AA{} and K-Cor images. Blue arrows indicate corresponding discernible features observed across images from different instruments.
\label{subimage}}
\end{figure*}

\begin{figure*}[ht!]
\centering
\includegraphics[width=1\textwidth]{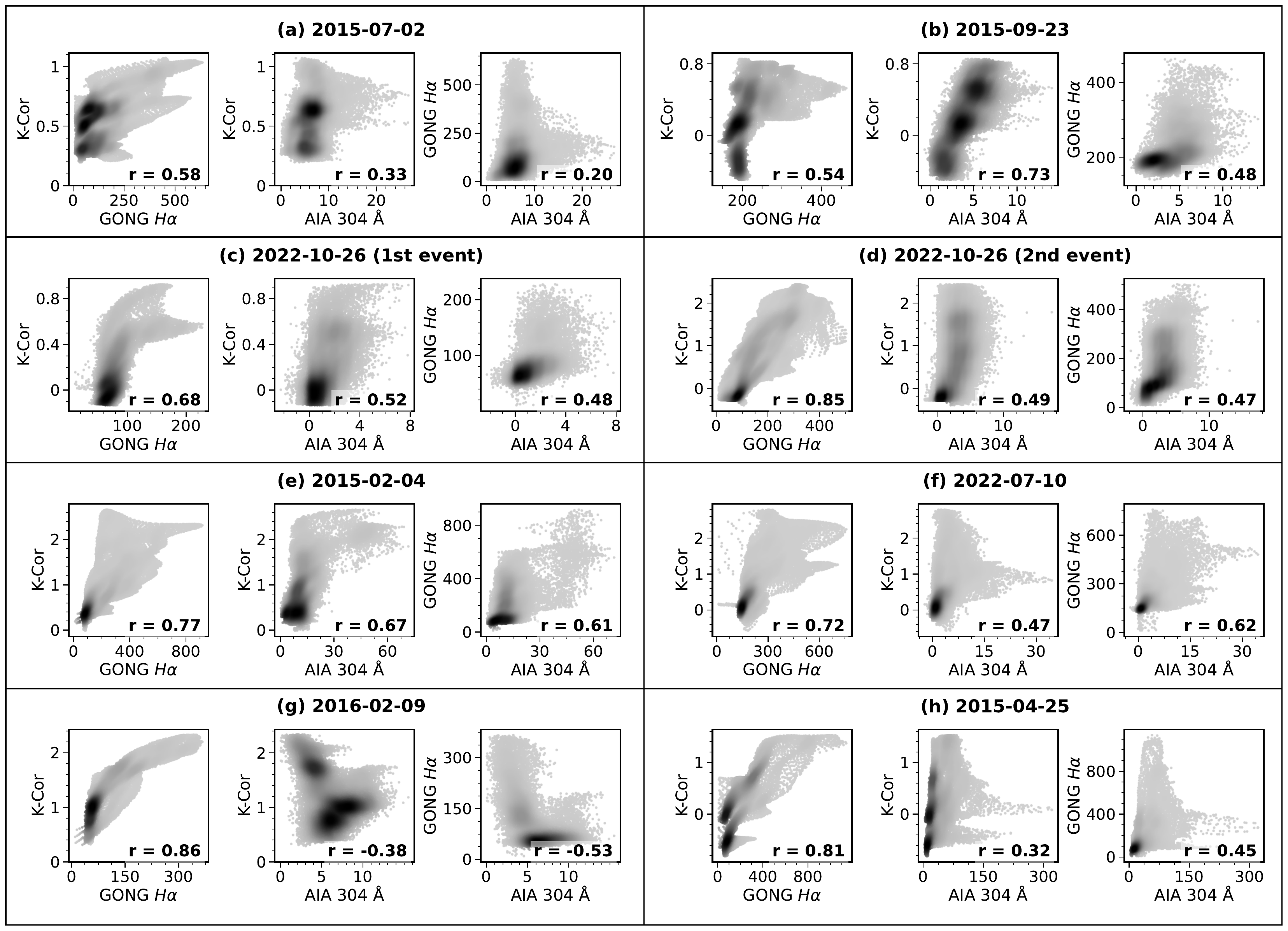}  
\caption{Quantitative comparison of CME core and prominence intensities using a combination of scatter plot and density distribution for three pairwise combinations: K-Cor vs. GONG H$\alpha$, K-Cor vs. AIA 304~\AA{}, and H$\alpha$ vs. AIA 304~\AA{}. Intensities (pixel values) from all three observations are shown in arbitrary units. The intensity comparisons are restricted to the ROIs shown in Figure~\ref{subimage}. The darker shade of black indicates a higher density of occurrence, whereas the lighter gray represents a low density distribution. For each of the eight events (four Category-A and four Category-B CMEs), the Pearson correlation coefficient ($r$) is calculated from the corresponding ROIs, with the values displayed in the bottom-right corner of each plot.
\label{ccplot}
}
\end{figure*}

\begin{figure*}[ht!]
\centering
\includegraphics[width=1\textwidth]{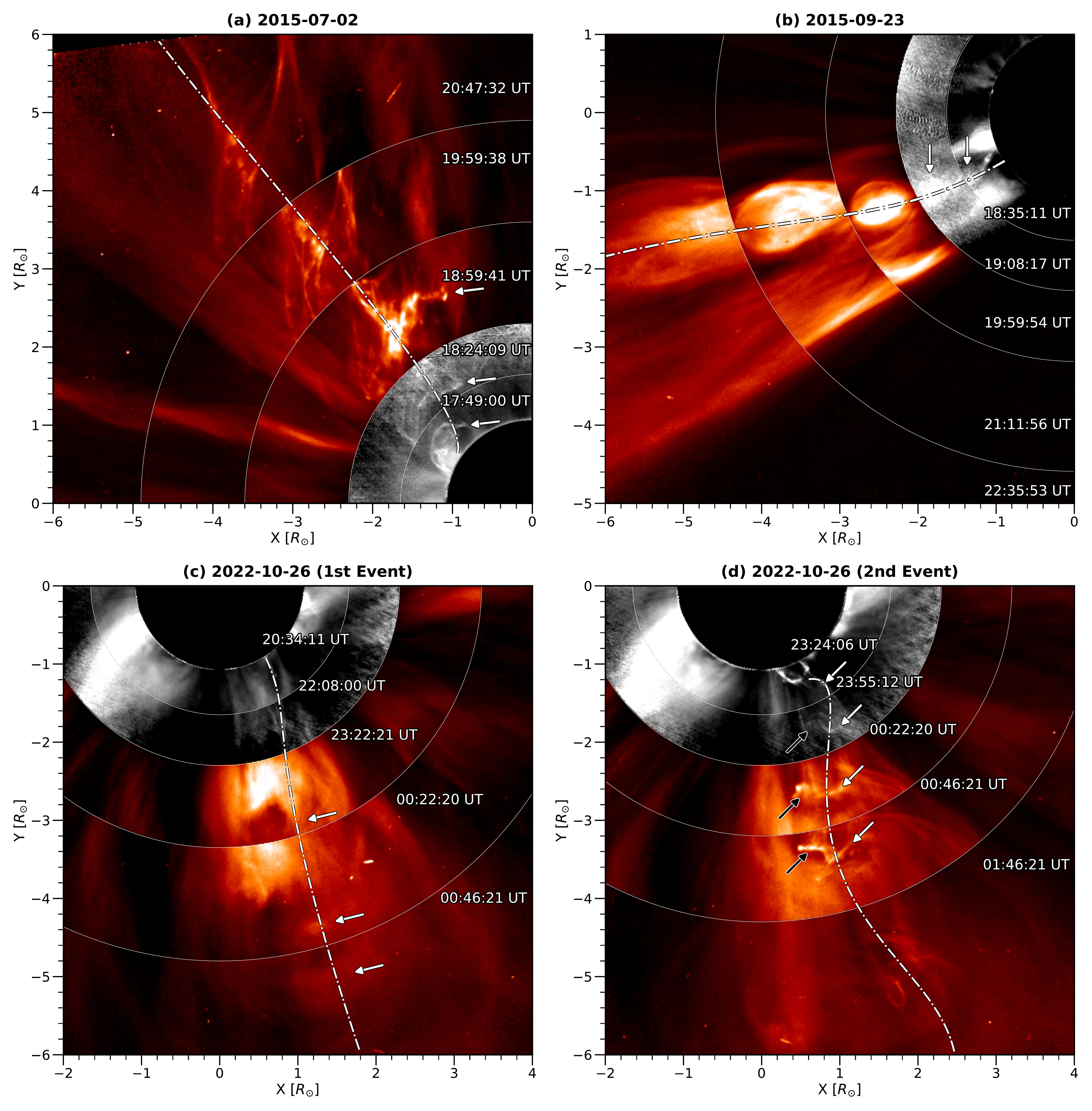}  
\caption{Composite images showing the temporal evolution of Category-A CMEs as they propagate from the inner corona (observed by K-Cor) into the outer corona (observed by LASCO/C2). Each panel corresponds to a distinct event: (a) 2015 July 02, (b) 2015 September 23, (c) 2022 October 26 (1st event), and (d) 2022 October 26 (2nd event). The K-Cor FOV spans from 1.05 to 2.3 $R_\odot$, beyond which LASCO/C2 observations begin. For each event, five time steps are shown in a single image, separated radially using segmented annular slices to illustrate the CME’s outward evolution (the first two from K-Cor and the remaining three from LASCO/C2). In each panel, a white dash–dot line traces the smoothed trajectory of the CME core’s apex, derived from a polynomial fit to the original path to highlight its motion through the corona. Panels (a), (b), and (c) are fitted with third-order polynomials, while panel (d) uses a fourth-order fit to capture its more complex evolution. The arrows illustrate how different CME core features evolve over time in the sequence of frames. An animation of this figure is available \href{https://drive.google.com/file/d/1ueYOCxYVJKXfWZIRyPC-u-X3QgNs1zhz/view?usp=drive_link}{online}.
\label{kcorc2}}
\end{figure*}

\begin{figure*}[ht!]
\centering
\includegraphics[width=1\textwidth]{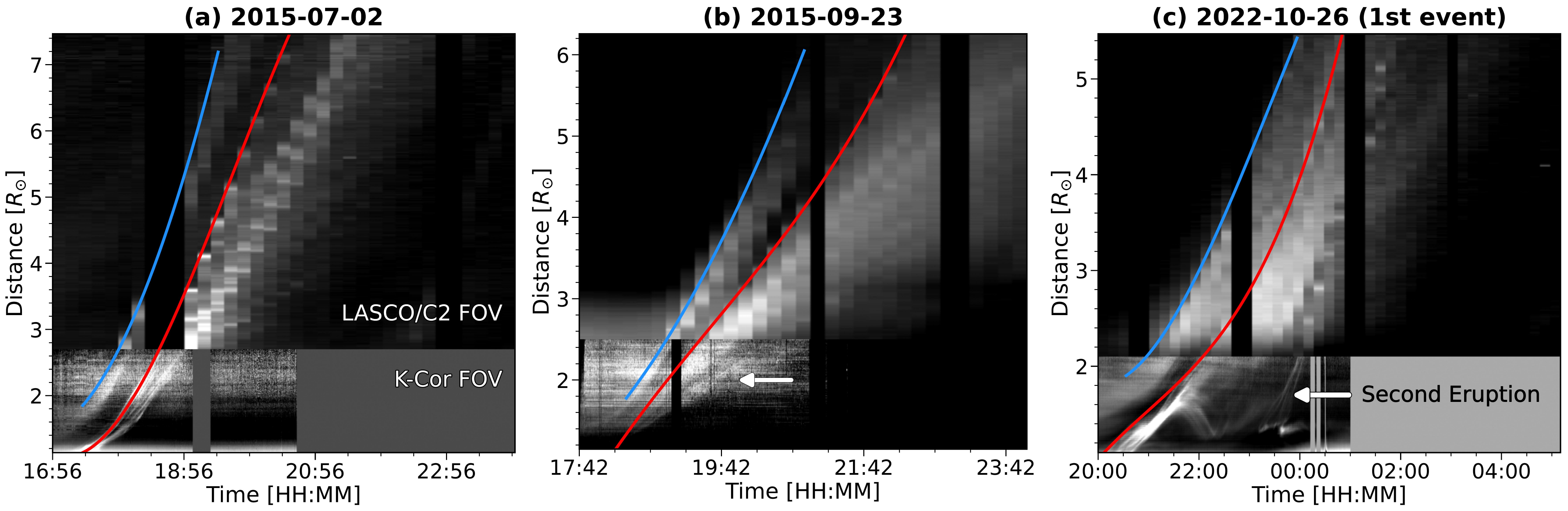}  
\caption{Distance–time maps illustrating the temporal evolution of CME core features along the traced trajectory in K-Cor and LASCO/C2 FOV as shown in Figure~\ref{kcorc2} (white dash-dot line) for three Category-A events: (a) 2015 July 02, (b) 2015 September 23, and (c) 2022 October 26 (1st event).
In each panel, the blue curve represents the manually traced path of the leading edge, while the red curve, obtained by polynomial fitting of the core trajectory shown in Figure~\ref{kcorc2}, indicates the path of the apex of the CME core.
The arrow in panel (b) indicates the evolution of the foot-point associated with the event, as discussed in more detail in Section~\ref{sec:res2}.
The arrow in panel (c) shows the evolution of the second eruption corresponding to the illustrated event.
\label{tvmap}}

\end{figure*}

\section{Analysis and Results} \label{sec:res}

\begin{deluxetable*}{clccccccc}
\tablewidth{0pt}
\tablecaption{Analysis results of Category-A CME events \label{tab:taba}}
\tablehead{
\colhead{S.No} & \colhead{Date} & \colhead{Eruption} &
\multicolumn{3}{c}{Correlation Coefficients (r)} &
\colhead{$\Delta\theta$ (Deg)} & \colhead{$\Delta T$ (min)} \\
\colhead{} & \colhead{(YYYY-MM-DD)} & \colhead{Time (UT)} &
\colhead{K-Cor vs H$\alpha$} & \colhead{K-Cor vs 304} & \colhead{H$\alpha$ vs 304} &
\colhead{} & \colhead{} & \colhead{}
}
\startdata
1  & 2014-05-28 & 17:14 & 0.81 & 0.65 & 0.76 & 15.41 & 92.18\\
2  & 2014-12-21$^\dagger$ & 01:52 & 0.67 & 0.27 & 0.30 & 6.33  & -10.40 \\
3  & 2015-05-05$^\#$ & 22:09 & 0.65 & 0.67 & 0.77 & 15.11 & -6.35  & \\
4  & 2015-05-15 & 21:28 & 0.68 & -0.22 & 0.11 & 11.28 & 31.35\\
5  & 2015-05-25 & 21:10 & 0.68 & 0.09 & 0.01 & 19.44 & 97.98\\
6  & 2015-07-02 & 17:12 & 0.58 & 0.33 & 0.20 & 17.01 & 35.28\\
7  & 2015-09-23$^*$ & 18:00 & 0.54 & 0.73 & 0.48 & 39.82 & 16.15 \\
8  & 2016-02-08$^*$ & 22:20 & 0.76 & 0.73 & 0.68 & 33.32 & 27.00\\
9  & 2016-08-08$^*$ & 20:22 & 0.80 & — & — & 14.79 & 138.35\\
10 & 2022-01-31 & 23:38 & 0.73 & 0.00 & 0.24 & 12.38 & 12.65 \\
11 & 2022-05-14$^{\#}$ & 17:09 & 0.84 & 0.62 & 0.53 & 17.05 & 92.83\\
12 & 2022-05-24$^\dagger$ & 22:19 & 0.58 & -0.03 & 0.02 & 19.39 & 4.38 \\
13 & 2022-09-24 & 19:57 & 0.63 & 0.48 & 0.36 & 6.54  & 80.01\\
14 & 2022-10-26$^{\dagger}$ & 20:00 & 0.68 & 0.52 & 0.48 & 5.70  & 103.50\\
15 & 2022-10-26$^\dagger$ & 21:28 & 0.85 & 0.49 & 0.47 & 30.25 & 16.60 \\
\hline
\textbf{Avg.} &  &  & \textbf{0.70} & \textbf{0.38} & \textbf{0.38} & \textbf{17.59} & \textbf{48.77} & \\
\enddata
\tablecomments{The table summarizes the parameters of Category-A CME eruptions that preserve their three-part structure into the LASCO/C2 field of view. The dates and times are taken from the K-Cor database. The table includes the Pearson correlation coefficients ($r$) between K-Cor, H$\alpha$, and 304 \AA{} ROI images, as well as the angular ($\Delta\theta$) and temporal ($\Delta T$) deviations for all Category-A CME events. The deviations represent the differences between the actual eruption position angle and onset time, and the values extrapolated from LASCO/C2 observations alone, in the absence of inner coronal coverage.
‘—’ marks entries where the value of $r$ could not be determined owing to the lack of corresponding AIA 304~\AA{} observations.}
\end{deluxetable*}

\begin{deluxetable*}{clcccccc}
\tablewidth{0pt}
\tablecaption{Analysis results of Category-B CME events \label{tab:tabb}}
\tablehead{
\colhead{S.No} & \colhead{Date} & \colhead{Eruption} &
\multicolumn{3}{c}{Correlation Coefficients (r)} \\
\colhead{} & \colhead{(YYYY-MM-DD)} & \colhead{Time (UT)} &
\colhead{K-Cor vs H$\alpha$} & \colhead{K-Cor vs 304} & \colhead{H$\alpha$ vs 304} & \colhead{}
}
\startdata
1  & 2014-02-20 & 22:24 & 0.70 & 0.45 & 0.31 \\
2  & 2014-04-29 & 19:40 & 0.62 & 0.37 & 0.40 \\
3  & 2014-06-30 & 17:33 & 0.56 & 0.56 & 0.41 \\
4  & 2014-12-13 & 21:47 & 0.55 & 0.66 & 0.54 \\
5  & 2014-12-21$^\dagger$ & 00:48 & 0.75 & 0.46 & 0.47 \\
6  & 2015-02-04 & 19:28 & 0.77 & 0.67 & 0.61 \\
7  & 2015-02-08 & 22:19 & 0.91 & 0.73 & 0.66 \\
8  & 2015-04-25 & 18:03 & 0.81 & 0.32 & 0.45 \\
9  & 2015-05-01 & 21:56 & 0.86 & 0.67 & 0.72 \\
10 & 2015-08-01 & 20:00 & 0.75 & 0.16 & 0.37 \\
11 & 2015-12-17 & 19:22 & 0.50 & 0.42 & 0.39 \\
12 & 2016-02-09 & 17:58 & 0.86 & -0.38 & -0.53 \\
13 & 2016-06-11 & 22:06 & 0.49 & 0.59 & 0.46 \\
14 & 2021-04-29 & 17:01 & 0.81 & 0.77 & 0.60 \\
15 & 2021-06-25 & 20:17 & 0.64 & 0.52 & 0.71 \\
16 & 2021-10-10 & 22:35 & 0.72 & 0.57 & 0.27 \\
17 & 2022-02-01 & 23:00 & 0.89 & 0.74 & 0.70 \\
18 & 2022-04-27 & 18:06 & 0.80 & 0.45 & 0.53 \\
19 & 2022-05-24 & 21:55 & 0.67 & 0.64 & 0.53 \\
20 & 2022-07-10 & 16:58 & 0.72 & 0.47 & 0.62 \\
21 & 2022-07-20 & 21:16 & 0.88   & — & — \\
22 & 2022-07-31 & 22:36 & 0.53 & 0.39 & 0.34 \\
23 & 2022-11-25 & 21:28 & 0.65 & 0.49 & 0.59 \\
\hline
\textbf{Avg.} &  &  & \textbf{0.72} & \textbf{0.49} & \textbf{0.46}   & \\
\enddata
\tablecomments{Same parameters as in Table~\ref{tab:taba}, but for Category-B CME events. Angular ($\Delta\theta$) and temporal ($\Delta T$) deviations are not included, as these events lacked a traceable three-part structure in LASCO/C2 observations, preventing reliable extrapolation.}
\end{deluxetable*}

\subsection{Comparison Between CME Core and Prominence in the Lower Corona} \label{sec:res1}

We examined the spatial and temporal correspondence between CME core structures observed in K-Cor white-light images and their associated prominence features observed in H$\alpha$ and AIA 304~\AA{} in the lower corona.
Figures~\ref{2july15sept} and \ref{26oct} showcase four Category-A CME events that retain their three-part structure into the LASCO/C2 FOV, selected to highlight the diversity in CME core evolution and morphology encountered in our study.

Figure~\ref{2july15sept}(a) depicts the event from 2015 July 2, showing the evolution of the CME across four time-steps in K-Cor white-light, H$\alpha$ and AIA 304~\AA{}.
The event can be characterized by the appearance of a bright core-like structure, a surrounding dark cavity, and a diffused leading edge in K-Cor white-light images.
The CME core structure aligns with the eruption of a prominence seen in both AIA 304~\AA{} and GONG H$\alpha$ images.
The spatial structures of the CME core seen in the K-Cor closely resemble the prominence seen in the H$\alpha$ as well as that in AIA 304~\AA{} observations.
Initially, the full prominence is visible in all three observations, suggesting a face-on orientation.  
As the event progresses, the prominence exits the H$\alpha$ FOV; however, its two foot-points remain discernible even in the later frames.
Overall, the spatial structure and evolution of the CME core shows coherence with that of the erupting prominence.
To facilitate a visual comparison between the CME core and the associated prominence structure, we have overlaid H$\alpha$ intensity contours (red) on K-Cor, AIA 304~\AA{} images.
Similarly, the front of the CME core traced from K-Cor data (blue contour) is overlaid on AIA 304~\AA{} images for visual guidance to follow its evolution.

Figures~\ref{2july15sept}(b), \ref{26oct}(a) present prominences with edge-on orientation, whereas Figure~\ref{26oct}(b) displays a face‑on prominence.
Figure~\ref{2july15sept}(b) corresponds to a CME event of 2015 September 23.
A three-part CME structure is clearly visible in the K-Cor observations, and the core of the CME is well aligned with the structure of the erupting prominence present in the H$\alpha$ and AIA 304~\AA{} images.

Figures~\ref{26oct}(a) and \ref{26oct}(b) depict two eruptions on 2022 October 26 that subsequently interacted to produce what appears to be a dual‑core CME (see Figure~\ref{kcorc2}(c) and \ref{kcorc2}(d)).
%
The first eruption shows a three-part structure in K-Cor images, where the core has an elongated structure matching the prominence observed in the H$\alpha$ and AIA 304~\AA{} images.
The animation corresponding to Figure~\ref{26oct} shows that part of the prominence material falls back to the solar surface, which is visible in AIA 304~\AA{} as well as in K-Cor observations.
The eruption, as mentioned earlier, is followed by another eruption where the corresponding prominence has a face-on orientation, which has a spatial and temporal coherent structure in the CME core structure observed in K-Cor.
One notable point is that the second eruption on 2022 October 26 does not exhibit a classical three‑part structure within the K‑Cor FOV.
This is most likely because the earlier eruption had already cleared the coronal plasma in that region, leaving no material to form a distinct leading edge.
As the two eruptions ultimately merge into a single CME that does show a three‑part structure in LASCO/C2 FOV, we include these events in Category-A.

A similar analysis was carried out for the Category-B CMEs, which reveal the three-part structure in K-Cor FOV but do not maintain it in the LASCO/C2 FOV.
Figure~\ref{4x4prominence} shows four such events at a single time stamp, allowing for comparison between the CME cores and their associated prominence structures.
The distinct core structures in K-Cor, along with their three-part configuration, are clearly visible and well-matched with the corresponding prominence features in both H$\alpha$ and 304~\AA{} images.
Animations of these events strongly suggest that the core part of CMEs evolves in tandem with the prominences observed in the AIA 304~\AA{} channel.

In all Category-A CMEs and Category-B CMEs, visual inspection reveals a strong spatial correlation between prominence structures observed in H$\alpha$, AIA 304~\AA{}, and CME cores observed in K-Cor images.
The structural boundaries in AIA 304~\AA{} and K-Cor, in particular, exhibit notably strong alignment within AIA/SDO FOV.

The above-presented analysis provides clear evidence that the CME core part has overall spatial and temporal correspondence with prominences.
To perform a more detailed spatial comparison, examining the presence of finer scale structure of prominences observed in H$\alpha$ and AIA~304~\AA{} and that in the K-Cor observations, we have focused on a common small region of interest (ROI).
The ROI was selected for each event, corresponding to the time when the prominence was most clearly visible across all three instruments (see yellow boxes in Figure~\ref{2july15sept}, \ref{26oct}, and \ref{4x4prominence}).
Figure~\ref{subimage} shows the zoomed-in views of the selected ROIs from GONG H$\alpha$, K-Cor, and AIA 304~\AA{} images. Panels (a–d) correspond to Category-A CMEs, while panels (e–h) display examples from Category-B CMEs.

Panels (a) and (d) show ROI highlighting the fine-scale structure of prominences and corresponding K-Cor counterparts observed on 2015 July 2 and the 2nd CME event on 2022 October 26, respectively.
Both the prominences have a face-on orientation.
The K-Cor image, consisting of the core part of the CME presented in panel (a), has structural similarity on a finer scale with the details present in the prominence seen in H$\alpha$.
For instance, a Y-shaped bright structure is visible in both the H$\alpha$ and K-Cor image, which is marked by the arrow in panel (a).
Interestingly, in the AIA 304~\AA{} image, the same structure appears as a dark feature, showing a contrasting characteristic.
Despite these subtle differences, the overall morphology of the prominence seen in AIA 304~\AA{} is consistent with that of the H$\alpha$ observations and CME core structure in the K-Cor image.
The prominence presented in panel (d) had an arc-like structure in the H$\alpha$ image, which shows remarkable similarity with the CME core structure seen in K-Cor data. There is a wide dark void at the apex of the prominence (see arrow in panel (d)) that is also present in the corresponding K-Cor image while being absent in AIA 304 \AA{}.

The prominence associated with the CME eruption on 2022 July 10 shows a tilted U-shaped structure (Figure~\ref{subimage}(f)) in the H$\alpha$ observations, and the corresponding CME core, as seen in K-Cor, has very similar morphology.
The AIA~304~\AA{} channel also exhibits very similar morphology.
The H$\alpha$ and AIA~304~\AA{} images in panel (h) display $\rm{\lambda}$-shaped structure in the observed prominence as well as in the K-Cor image representing the core structure of the CME.
The prominences presented in panels (b), (c), and (e) have edge-on orientations and display elongated structure.
The $H\alpha$ and AIA~304~\AA{} images show similar morphological structure as present in the K-Cor observations.
Panel (g), however, shows a case similar to panels (a) and (d), where the AIA 304~\AA{} image reveals a structure that contrasts strongly with those in H$\alpha$ and K-Cor. At the location where H$\alpha$ and K-Cor display an elongated bright feature, AIA 304~\AA{} instead shows a dark structure (see arrow in panel (g)).

Overall, we see that there is a remarkable similarity on a finer scale between prominence structures observed in H$\alpha$ observations and the core part of the CMEs observed in the K-Cor observations. The AIA~304~\AA{} observations also show prominence structure consistent with that of H$\alpha$.
However, in many events, AIA~304~\AA{} displays structures on a finer scale which do not match with H$\alpha$ and K-Cor.
Overall, a common trend observed in both Category-A and Category-B CMEs is that the K-Cor core shows a stronger visual correspondence with the H$\alpha$ prominence structures compared to AIA 304~\AA{}.

One notable observation is that the GONG H$\alpha$ shows sharper structure compared to K-Cor.
Additionally, when comparing H$\alpha$ and AIA 304~\AA{} images, an interesting feature is that the H$\alpha$ prominence structures tend to appear narrower than their counterparts in AIA 304~\AA{}, as most evident in panels (b), (c), and (e)—a trend also reported for on-disk filaments \citep[e.g.][]{Vial_2012}.

To gain a clearer understanding of the one-to-one structural correspondence with a quantitative analysis, an image correlation analysis was carried out between each pair of instruments—namely, K-Cor vs. H$\alpha$, K-Cor vs. AIA 304~\AA{}, and H$\alpha$ vs. AIA 304~\AA{}.
The image correlation was carried out only on a small portion of the common FOV among all three instruments, for example, ROIs displayed in Figure~\ref{subimage}.
For this analysis, the K-Cor and H$\alpha$ images within the selected FOV were upscaled to match the pixel size of the AIA images. After ensuring identical pixel size for all three instruments, Pearson correlation coefficients ($r$) were computed between the pixel values of images of each instrument pair (see Figure~\ref{ccplot}, panels (a–d) for Category-A CMEs and panels (e–h) for Category-B CMEs).

For Category-A events, the average correlation between K-Cor and H$\alpha$ was approximately 0.70 while the correlations between K-Cor and AIA 304~\AA{}, and between H$\alpha$ and AIA 304~\AA{}, were notably lower, around 0.4 each (see Table~\ref{tab:taba}).
In the case of Category-B events, the average K-Cor–H$\alpha$ correlation was about 0.72, with correlations of around 0.49 between K-Cor and AIA 304~\AA{}, and 0.46 between H$\alpha$ and AIA 304~\AA{}.
These values are slightly higher than those for Category-A, though they remain within a similar range (see Table~\ref{tab:tabb}). Additionally, Category-B events displayed a broader spread in correlation values, likely due to the larger number of events in the sample. For instance, the event of 2016 February 9 (Figure~\ref{ccplot}(g)) shows a negative correlation in AIA 304~\AA{}, where the observed feature contrasts with that seen in the K-Cor and H$\alpha$ images.
It is also important to note that these correlation values were derived using upscaled K-Cor and H$\alpha$ images. When instead the AIA and H$\alpha$ images are downscaled to match the K-Cor resolution, the average correlations increase slightly—by approximately 0.03 for K-Cor–H$\alpha$ and by about 0.12 for K-Cor–AIA in Category-A events. Similar trends are also found for Category-B CMEs, with average increases of about 0.04 and 0.10, respectively. Nevertheless, the overall results consistently indicate that the CME core observed in K-Cor correlates more strongly with H$\alpha$ prominence features than with those observed in AIA 304~\AA{}.

Several minor ambiguities were also observed in the structural comparisons discussed above.
For example, in Figure~\ref{2july15sept}(a), additional features are visible in the CME core in K-Cor and in the prominence in AIA 304~\AA{} marked by white arrows, which are not apparent in the corresponding H$\alpha$ images.
These differences may be due to the Doppler dimming effect \citep{Peat_2024} inherent in narrowband H$\alpha$ observations.
Prominence plasma with significant LOS velocity can experience Doppler shifts that reduce the detected H$\alpha$ emission, in some cases rendering the feature entirely invisible in narrowband filters.
Additionally, some discrepancies may arise from the presence of hotter plasma structures in the prominence-corona transition region (PCTR), which are not captured in the H$\alpha$ line.
It is worth noting that similar effects may also explain the absence of identifiable prominence structures in H$\alpha$ for the 18 events eliminated from the original set of 65.
\subsection{Tracking CME core from lower to higher corona} \label{sec:res2}

Having already established the association between the core of their three-part structure and prominence structures in the inner corona, we aim to examine whether this relationship persists in the outer corona.
We investigated the evolution of 15 Category-A CMEs listed in Table~\ref{tab:taba} as they propagate from the K-Cor FOV into that of LASCO/C2.

The analysis presented in Section~\ref{sec:res1} is limited to the FOV of GONG H$\alpha$ and AIA 304~\AA{}, which extends only up to 1.13~R$_{\odot}$ and 1.3~R$_{\odot}$, respectively.
Due to the lack of narrowband EUV or H$\alpha$ observations beyond these heights and extending into the LASCO/C2 FOV, the present analysis relies on visually linking the CME core observed in K-Cor with its counterpart in LASCO/C2. 
This linkage is based on the established morphological relationship between the prominence structures and their associated CME core within 1.3~R$_{\odot}$, as observed with the K-Cor data.
We then track the apparent continuity of the CME core as it propagates outward.
The primary objective is to assess whether the core structure observed at the outer edge of the K-Cor FOV continues into the structure identified as the core in LASCO/C2.
This allows us to determine the morphological continuity of the CME core from the inner to the outer corona.

Figure~\ref{kcorc2}(a) illustrates the evolution of the CME event on 2015 July 2 using a composite of five radially segmented annular slices: the first two from K-Cor and the remaining three from LASCO/C2, each corresponding to a different time step.
These slices are arranged in a single image to effectively show the CME core’s progression through the corona.
Readers can view the corresponding animation to have a better visualization of the CME evolution.
In the innermost slice, the core appears as a structure with two foot-points anchored to the solar disk.
These foot-points correspond to prominence foot-points as seen in H$\alpha$ and AIA 304~\AA{} images presented in Figure~\ref{2july15sept}.
As the eruption progresses, the right foot-point (marked with an arrow) appears to disconnect from the disk and rise more rapidly than the left, which remains anchored.
In the second slice, the apex of the core appears to be present just beyond the selected slice. The disconnected right foot-point is also visible (marked with an arrow), while the central body of the structure appears blurred due to noise. Then, the LASCO/C2 slices can be seen capturing a clear central portion of the elongated core structure.
For effective visual tracking, a white dot–dashed curve is overlaid on the composite image, which is obtained simply from polynomial fitting to the trajectory of the apex part of the CME core.

Figure~\ref{kcorc2}(b) presents similar composite images for the event on 2015 September 23.
In the event, the CME core can be seen as a blob-like structure (shown by white arrows), with a clearly visible foot-point that evolves smoothly into the outer corona.
Figure~\ref{kcorc2}(c) and (d) present two successive events that occurred on  2022 October 26.
Both events are associated with two separate prominences that erupted sequentially.
As they evolve into the LASCO/C2 FOV, the two eruptive structures appear to merge and form the core of a single CME. As in Figure~\ref{kcorc2}(a), arrows are used to track and highlight the evolving core structures across the radial slices. 
See the accompanying animation to have a better understanding of the evolution of these events.

To further analyse the kinematic continuity of the CME core, we extracted intensity values along the trajectory traced by the white dot-dashed curves in Figure~\ref{kcorc2}(a)–(d) and created distance-time maps.
Figure~\ref{tvmap}(a) presents the distance-time map of the 2015 July 2 event. The distance axis begins at the height where tracking starts within the K-Cor FOV—typically near the lower boundary of the K-Cor observational range.
Although K-Cor formally extends to approximately 3 $R_\odot$, its outer regions are often affected by significant noise, as evident in the image.
Therefore, only the region with reliably identifiable structures was used in the analysis, followed by data from the LASCO/C2 FOV.
The red curve corresponds to a polynomial fit to the trajectory of the apex of the CME core, while the blue curve traces the trajectory of the apex of the CME leading edge.
A data gap appears in the LASCO/C2 just as the CME core enters its FOV, obstructing visual tracking in standard images.
However, the distance–time representation still clearly captures the continuous evolution of the CME core across the data gap.

Figure~\ref{tvmap}(b) shows the 2015 September 23 event.
But here, an ambiguity can be seen and explained as follows: As noted earlier, this event exhibits both a blob-like head structure and a foot-point on the plane of the sky. In the K-Cor–H$\alpha$ comparison (Section~\ref{sec:res1}), we focused on the foot-point structure, since the blob does not appear until about 1.5 R$_\odot$, outside the H$\alpha$ and AIA 304~\AA{} FOV.
In contrast, in this kinematic study, we track the blob-like head (shown by the red curve) rather than the foot-point. So, in summary, the ambiguity is that, in both analyses, we are not comparing the same feature.
Nevertheless, both the foot-point and the blob can be identified as extensions of the same structure.
Based on the on-disk eruption captured in AIA 304~\AA{}, we speculate that the blob could be formed from the LOS evolution of the second foot-point of the edge-on prominence, which likely became diffuse due to rapid expansion, with portions of the material possibly falling back to the solar surface.
In Figure~\ref{tvmap}(b), both the leading blob and the foot-point (marked by a white arrow) can be distinctly identified. 
Please refer to the accompanying animation of the event of 2015 September 23 corresponding to Figure~\ref{kcorc2} for a clearer understanding.
    
Figure~\ref{tvmap}(c) summarizes both eruptions from 2022 October 26 in a single map, constructed using pixel values extracted along the trajectory of the first core.
The first eruption, originating from an edge-on prominence, evolves into a core within the K-Cor FOV, but a significant portion of its structure appears to fall back to the Sun, as seen in Figure~\ref{tvmap}(c) from around 22:00 UT. 
This falling material subsequently merges with the second eruption.
Additionally, the first core shows apparent signs of rotation in the plane of the sky, contributing to a complex dynamical evolution. The leading edge of the first CME is prominent in the K-Cor images but appears to diffuse as it propagates into the LASCO/C2 FOV.
The second eruption followed a more irregular trajectory due to its interaction with the first eruption, as evident in Figure~\ref{kcorc2}(d), and appears as a narrow structure in Figure~\ref{tvmap}(c).

It is also important to highlight several challenges encountered in tracking the CME core, which can be outlined as follows: 
    
\begin{itemize}
    \item \textbf{Morphological challenge:} In some cases (e.g., events marked with $\#$ in Table~\ref{tab:taba}), a three-part structure is visible in LASCO/C2, but it is either weakly defined or not sustained throughout the full LASCO/C2 FOV. As a result, it becomes difficult to reliably identify and track the CME core to the outer edge of the LASCO/C2 observations.

    \item \textbf{Tracking mismatch:} In events marked with an $\ast$ in Table~\ref{tab:taba} (for example, 2015 September 23), the feature tracked from K-Cor to LASCO/C2 may not correspond exactly to the one analyzed in the K-Cor–H$\alpha$ image comparison (Section~\ref{sec:res1}), owing to the complex evolution within the K-Cor FOV. Nevertheless, it can be verified that the features compared in these different analyses originate from the same prominence structure.

    \item \textbf{Event interaction:} Events marked with $\dagger$ in Table~\ref{tab:taba} involve interactions between multiple erupting structures. In particular, the merging of two erupting cores (e.g., 2022 October 26) complicates the identification of the CME's true source and its evolution into a coherent core structure. 
\end{itemize}

In summary, despite the presence of some challenges, the analysis shows that for most Category-A CMEs, the core structure observed in K-Cor exhibits clear morphological continuity as it propagates into the LASCO/C2 FOV.
This continuity suggests that the CME core remains closely associated with the corresponding prominence even at greater heights in the outer corona.

We also investigated the implications of associating CMEs with on-disk events in the absence of inner coronal observations by backtracking the CME core observed in LASCO/C2 onto the solar disk under simplified assumptions.
While a CME may exhibit varying speeds while evolving, it is standard practice to assume constant velocity and linear propagation within the LASCO/C2 FOV for simplicity when estimating average velocity for a large statistical study \citep{Yashiro_2004, Gopalswamy_2009}. 
This assumption is supported by our findings, which show that CME cores typically follow a linear trajectory with very low acceleration in the LASCO/C2 FOV compared to the inner corona, consistent with earlier results \citep{Bien_2011}.
Based on this linear approximation, we assessed the spatial and temporal uncertainties in identifying the CME core’s source region using only its motion in the outer corona.
Spatially, this refers to potential offsets in position angle between the projected and actual source regions; temporally, it refers to differences between the extrapolated and actual eruption times.

\begin{figure*}[ht!]
\centering
\includegraphics[width=1\textwidth]{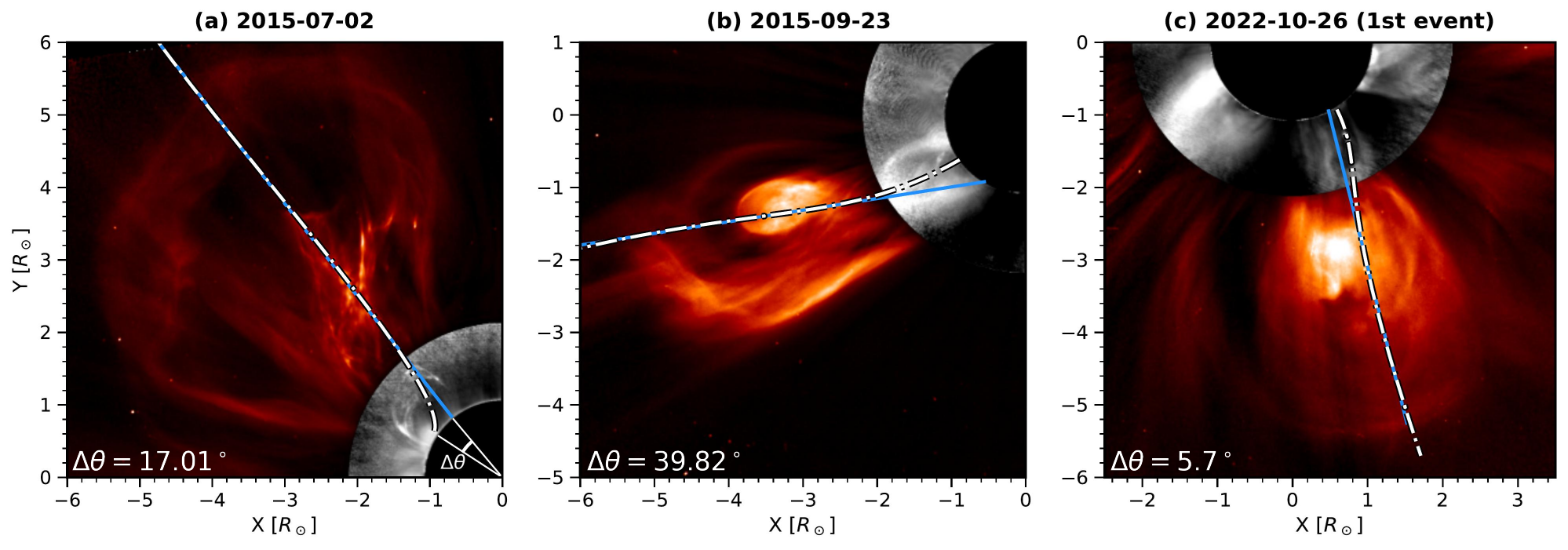} 
\caption{Assessment of spatial discrepancy in locating the position angle of CME core origins with only outer coronal (LASCO/C2) observations, without inner coronal data. For three Category-A CME events—(a) 2015 July 02, (b) 2015 September 23, and (c) 2022-10-26 (1st event)—the white dash-dot curve shows a third-order polynomial fit to the observed path of the CME core from the disk through K-Cor to LASCO/C2. The blue line shows a simple linear extrapolation of the trajectory based only on outer coronal (LASCO/C2) motion, projected back toward the solar surface. The angular deviation ($\Delta\theta$) between the observed and extrapolated paths is illustrated in the bottom-right corner of panel (a), with the corresponding values listed in the bottom-left corner of each panel.
\label{deltheta}}
\end{figure*}

To estimate the spatial discrepancy, we extrapolated the CME core’s trajectory (marked in white dot-dashed curve) observed in LASCO/C2 back to the solar disk, as illustrated in Figure~\ref{deltheta}, where the blue line indicates the extrapolated path.
The angular deviation ($\Delta \theta$) between the extrapolated location and the actual source region was then calculated. For this, we identified the associated prominence in AIA 304~\AA{} and used its midpoint as the reference point for the position angle of CME origin.
The position angle corresponding to the extrapolated path was compared to this reference to determine $\Delta \theta$. 
In Figure~\ref{deltheta}(a)–(c), $\Delta \theta$ is displayed in the lower left corner. A full list of $\Delta \theta$ values for all events can be found in Table~\ref{tab:taba}, with an average angular deviation of approximately 17.5\textdegree\, and with a maximum deviation of about 40\textdegree.
It should be noted that prominences can be extended and structurally complex, and their spatial extent can sometimes be comparable to the value of $\Delta \theta$.

\begin{figure*}[ht!]
\centering
\includegraphics[width=1\textwidth]{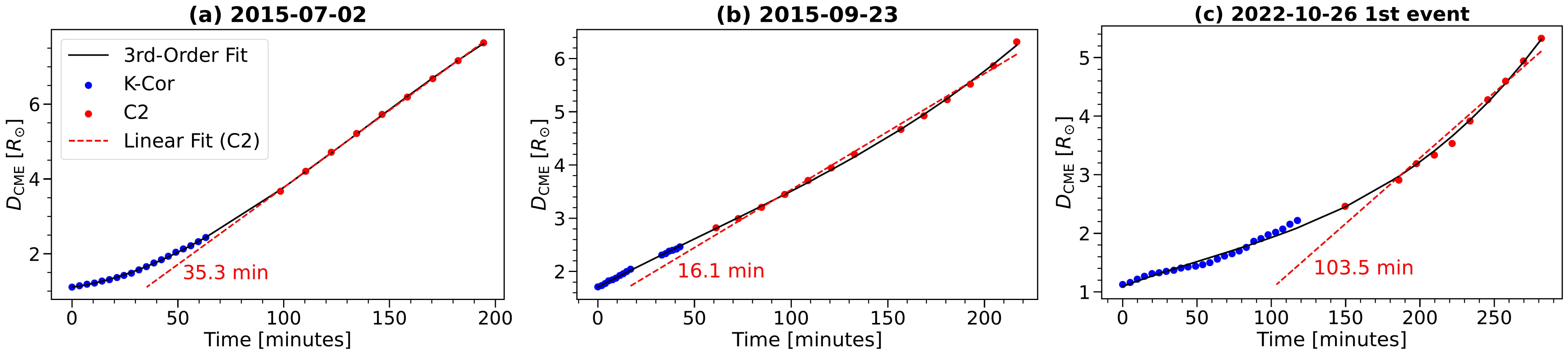} 
\caption{Assessment of temporal discrepancy in identifying CME core origin times when relying solely on outer coronal (LASCO/C2) observations. For each of the three Category-A events—(a) 2015 July 02, (b) 2015 September 23, and (c) 2022 October 26 (1st event)—a distance ($D_{\mathrm{CME}}$) vs time plot is shown along the actual CME core trajectory. Blue points mark positions tracked in K-Cor images, while red points correspond to LASCO/C2 detections. A third-order polynomial fit to all data points (black solid line) provides a smooth kinematic profile. A linear fit to LASCO/C2 points only (red dotted line) extrapolates the core motion backwards in time, assuming constant velocity. The temporal offset between the extrapolated and actual origin times ($\Delta T$) is indicated in each panel.
\label{deltaT}}
\end{figure*}

To assess temporal discrepancies, we constructed distance–time plots tracking the motion of the CME core apex using points extracted from both K-Cor and LASCO/C2 fields of view.
A third-order polynomial fit (black curve) was applied to the combined dataset, as shown in Figure~\ref{deltaT}.
Additionally, assuming a constant CME speed in the LASCO/C2 FOV, we performed a linear extrapolation using only LASCO/C2 data points (red dashed curve) to estimate the expected initiation time at lower coronal heights—mimicking a scenario where only LASCO/C2 observations are available.
Higher-order polynomial fits were not applied to the LASCO/C2 data, since we assumed a constant velocity. Although such fits can effectively trace the observed trajectory, they often overfit and may yield unreliable extrapolations due to overshooting.

\begin{figure}[ht!]
\centering
\includegraphics[width=\columnwidth]{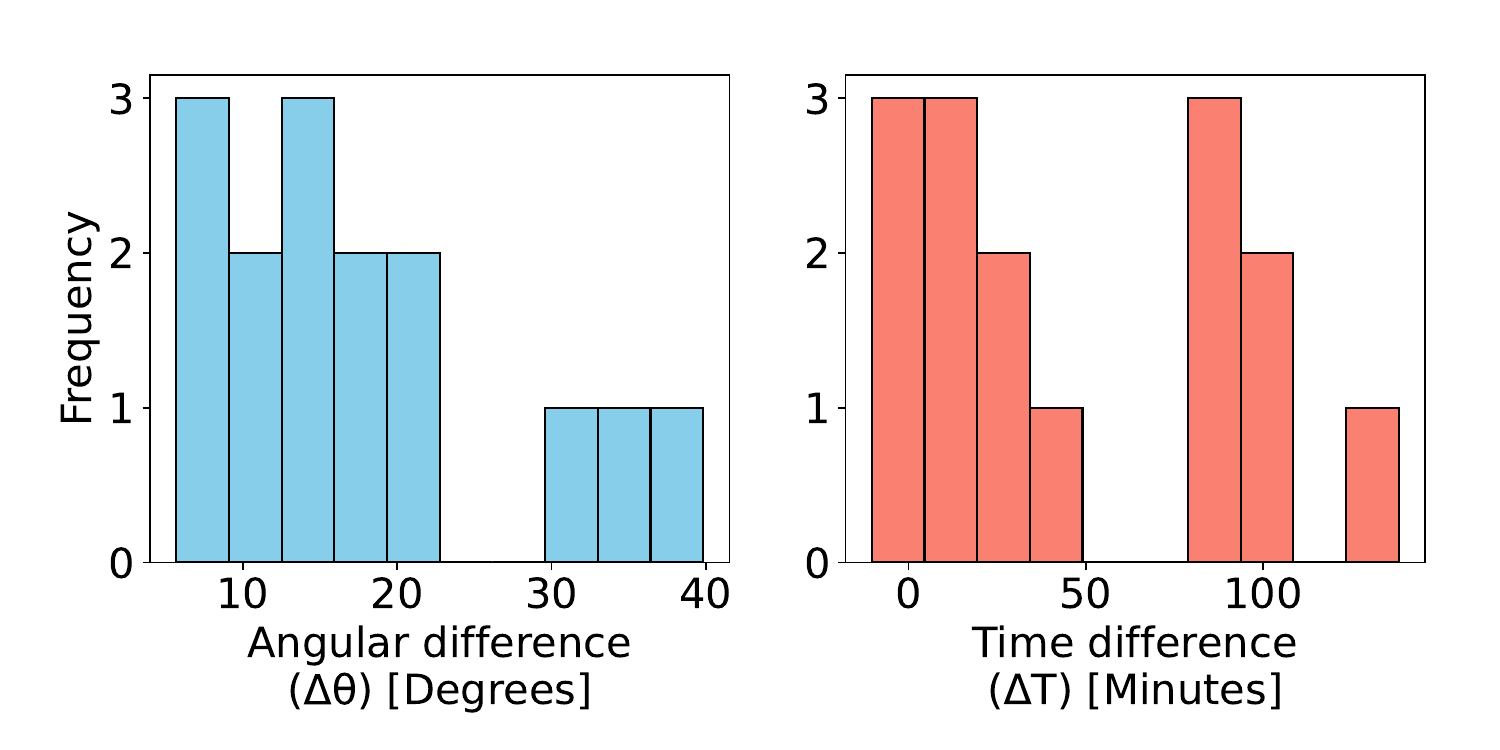} 
\caption{The left panel shows the histogram of angular or spatial offsets ($\Delta \theta$), representing the differences between the backtracked position angles of CME cores (from LASCO/C2 observations) and the actual position angles of the associated prominences observed in AIA 304~\AA{} for 15 Category-A CME events.
Similarly, the right panel shows the histogram of temporal offsets ($\Delta T$), comparing the backtracked onset times of the CME cores with the observed onset times of the associated prominences.
\label{deltahisto}}
\end{figure}

The temporal discrepancy ($\Delta T$) was then defined as the difference between this extrapolated onset time and the actual observed start time of core tracking. In Figure~\ref{deltaT}, the $\Delta T$ values are indicated in red for each event.
It is important to emphasize that this start time does not refer to the initial appearance of the prominence, but rather to the point at which the CME core could be reliably identified and tracked.
The $\Delta T$ values for all analyzed events are summarized in Table~\ref{tab:taba}, with an average temporal discrepancy of approximately 49 minutes and with a maximum discrepancy of about 138 minutes.
For a clearer representation of these discrepancies, Figure~\ref{deltahisto} presents histograms of both $\Delta \theta$ and $\Delta T$.
These deviations, although moderate, suggest that associating CME cores with their disk counterparts based solely on outer corona observations may lead to spatial and temporal misidentifications.

\section{Discussion} \label{sec:dis}

We investigated the connection between prominences and CME cores in the plane of the sky, using simultaneous observations in H$\alpha$, EUV 304~\AA{}, and white-light coronagraph data for a large sample of limb events.
Most of the analyzed CME events are a subset of events studied by \citet{Song_2023_1}. 
In this study, we have only focused on CMEs associated with the clear prominence eruption visible in the GONG H$\alpha$ and AIA~304~\AA{} observations, thus assessing the nature of core of CMEs only for prominence eruption related events. 
We have analyzed 38 events selected based on the criteria mentioned in Section~\ref{sec:data2}.
In agreement with \citet{Song_2023_1}, we confirm that all of our events show the three-part structure in the K-Cor FOV, but only 15 events (39$\%$ of the analyzed events) retain it in the LASCO/C2 FOV. 

In all the events, the core part of CMEs is visible in the K-Cor FOV in the very initial stages of eruption or sometimes prior to the onset of the CME eruptions. 
This enabled a direct comparison of CME cores with their associated prominences using simultaneous observations in white light, H$\alpha$, and 304 \AA{} within the height range of 1.05–1.13~R$_{\odot}$.
To explore the connection between CME core and prominence beyond 1.13~R$_{\odot}$, we looked at the AIA 304~\AA{} channel, which provided chromospheric context up to 1.3~R$_{\odot}$.

We found a remarkable similarity between the CME core structures and the prominence features observed in H$\alpha$.
For example, in prominences with a face-on orientation, the K-Cor observations reveal a similar pattern in the CME cores.
Similarly, in prominences with an edge-on orientation, the corresponding CME cores exhibit an elongated morphology.
Not only does the overall morphology of the prominences match the structures of the CME cores, but a detailed comparison also shows that the fine-scale substructures within the prominences are strikingly well reflected in the substructures of the CME cores.
Our quantitative comparison, carried out over ROIs covering the prominence and its surroundings, shows that H$\alpha$ images and K-Cor images have an average cross-correlation coefficient of $\sim0.7$.
Apart from the striking correspondence between CME core structures and prominence features, we also find that these core structures remain clearly traceable as they evolve beyond 1.13~R$_{\odot}$ within the K-Cor FOV. 
This extended traceability highlights not only the close relationship between the CME core and erupting prominence in the inner corona, but also the continuity of these features as the CME propagates outward.

We also performed image cross-correlation between the K-Cor and AIA~304~\AA{} images, for the same ROIs for which cross-correlation was estimated for the K-Cor and H$\alpha$ images. 
We find that on average the cross-correlation is $\sim$0.5, which is weaker compared to that for H$\alpha$.
One possible explanation for the weaker cross-correlation is that the AIA~304~\AA{} channel has the maximum response at  $10^{4.7}$~K; however, it also has a response at coronal temperature \citep{boerner_2012,AIA_2012}.
Therefore, ~304~\AA{} not only traces the cooler prominence plasma, but also the surrounding hotter PCTR plasma.
Our results indicate that in the inner corona, the white-light emission from the CME core shows a stronger correlation with the H$\alpha$ plasma, which probes the prominence core, rather than with the multi-thermal plasma observed in the AIA 304~\AA{}.
This is consistent with several previous studies, as summarized in the review by \citet{Labrosse_2010}, which report a broad range of electron densities in prominences. 
These studies indicate that electron densities are significantly higher in the cool core compared to the surrounding PCTR, thereby enhancing the Thomson scattering signal observed in white-light coronagraph images.

We further examined whether any strong chromospheric emission lines fall within the K-Cor bandpass (7200–7500~\AA{}) that could significantly contribute to the visibility of cool prominence plasma in white light.
Two \ion{O}{2} lines, at 7320.94~\AA{} and 7332.76~\AA{} \citep{Rivera_2019}, are present within the K-Cor bandpass.
These lines have a temperature response peaking around $10^{4.45}$~K, within the temperature range of prominence plasma.
However, the analyzed quantity from K-Cor is the polarization brightness and not the total brightness; therefore, the only way these two \ion{O}{2} lines can contribute to the observed core structure is through resonant scattering polarization. 
It requires detailed modeling effort to know if there is any significant contribution from the \ion{O}{2} lines in the measured linear polarization and its comparative contribution to the continuum polarization within the K-Cor passband, which is out of scope for the current work. 
However, since the K-Cor passband is 300~\AA{} wide we anticipate that continuum polarization will dominate over resonant scattering polarization from the \ion{O}{2} lines, if any.

Our observations do not have chromospheric counterparts beyond the 1.3~R$_{\odot}$, therefore we can not directly access the presence of cool prominence plasma at higher radial distances. 
However, to explore indirect evidence, we have tracked CME cores observed in K-Cor images and followed their evolution in LASCO/C2 observations. 
To track CME core at larger distances, we have first identified features of a CME core which show clear matching with prominence observed in the H$\alpha$ and 
AIA 304~\AA{} observations, then we followed these features transiting from the K-Cor to the LASCO/C2. 
In 15 Category-A events, we find strong evidence that CME cores have clear counterparts in prominence structures, and their kinematic evolution remains continuous into the LASCO/C2 FOV.
In many cases, we were able to track spatially resolved features from the inner corona up to $\sim$6~R$_{\odot}$, suggesting the CME cores consist of prominences. 

Our results are also consistent with a recent study by \cite{Mierla_2024}, they have directly observed the prominence and CME core in both white-light and EUV using multi-vantage-point observations to establish their correspondence.
They utilized the Full Sun Imager (FSI) on Solar Orbiter’s Extreme Ultraviolet Imager \citep[EUI,][]{Solo_eui_2020} in the 304~\AA{} channel to detect prominence material as far as 6~R$_\odot$, supported by white-light coronagraph data from LASCO/C2 and STEREO/SECCHI. 
Such observations are rare due to a lack of regular simultaneous observations in EUV and white light beyond 2~R$_\odot$. 
Aside from this case study, a recent statistical analysis by \citep{Pooja_2025} found that in 78\% of CMEs associated with erupting prominences, the CME cores show strong kinematic correspondence with their associated prominences. This result was obtained by comparing CMEs observed in the LASCO/C2 FOV with prominences seen in the AIA FOV for a large sample of 662 events.

The primary reason behind challenges to the idea of prominence core connection stems from the fact that not all CMEs are associated with prominence eruptions.
Therefore, alternate mechanisms are proposed to explain the origin of CME cores. For example, \citet{Howard_2017} proposed that the core represents the twisted central portion of an MFR, with the entire MFR appearing as the CME.
In their study, conducted without inner coronal observations, they traced CME structures from LASCO/C2 back to the solar disk and found significant spatial and temporal offsets between the close-by limb prominence and the core seen in coronagraph images.
Consequently, they suggested that even for CMEs associated with prominence eruptions, the cores are unlikely to be directly linked to prominences.
Based on their findings, \citet{Howard_2017} concluded that CME cores containing prominence material are “so rare that they are statistically insignificant.” 
    
In contrast, our study revisits this claim using a larger and more detailed dataset. 
By analyzing the spatial offset ($\Delta \theta$) between the CME core’s expected position angle and the associated prominence’s origin angle, as well as the temporal offset ($\Delta T$) between the CME core’s expected eruption time and the associated prominence eruption time, we find that both offsets can be significant.
Additionally, our observations reveal that in prominence-associated CMEs, it is common for the core to exhibit geometric and morphological changes as it evolves through the corona. 
These findings highlight the difficulty in establishing a reliable structural association between the prominence and the CME core in the absence of inner coronal observations. 

\cite{Song_2023_2} proposed a unified explanation for the nature of the three-part CME structure, in which the CME core primarily corresponds to an MFR, while any associated prominence represents additional material embedded within the MFR.
This interpretation builds upon a series of earlier studies \citep{Song_2017, Song_2019, Song_2019_1, song_2022}.
For prominence-associated CME eruptions, they suggest that the prominence appears brighter and more distinct, while the surrounding MFR remains faint and often
barely visible in coronagraph images.
For example, in some LASCO/C2 observations of three-part CMEs, they have identified the diffuse, fuzzy portion of the core as the MFR, whereas the sharper, filamentary features as embedded prominence material.
Our study, though focused on prominence-associated CMEs, does not provide direct evidence of MFR signatures in the core.
However, if the interpretation above holds, the signatures of MFR could be very faint and not discernible given the sensitivity of a ground-based coronagraph like K-Cor; thus, our results do not necessarily contradict this scenario.

Although our study provides insights into the nature of the CME cores using multi-wavelength observations, a complete understanding of its physical origin and its connection to the MFR requires direct magnetic field measurements with high temporal resolution, but such observations are currently lacking.
Instruments such as COronal Multi-channel Polarimeter \citep[CoMP,][]{Comp_2008} and Upgraded COronal Multi-channel Polarimeter \citep[UCoMP,][]{Ucomp_2016} provide valuable diagnostics of coronal magnetic fields, yet their utility has so far been limited to relatively stable structures like coronal cavities \citep{Bak_2013, Ruminiska_2022} and streamers \citep{Gibson_streamer_2017}, and they lack the sensitivity to capture highly dynamic phenomena such as CMEs.

\section{Conclusion} \label{sec:con}
In this work, we investigated the nature of CME cores by analyzing 38 prominence-associated CMEs using simultaneous H$\alpha$, AIA 304 \AA{}, and white-light coronagraph observations. 
All selected events exhibited the classical three-part structure within the K-Cor FOV, but only $\sim$39\% maintained it into the LASCO/C2 FOV. 
Our analysis reveals a clear one-to-one correspondence between prominence morphology and CME cores. 
Both face-on and edge-on prominence orientations are consistently reproduced in the cores, with fine-scale substructures also well preserved. 
A quantitative comparison further supports this connection. 
The average cross-correlation is strong between H$\alpha$ and K-Cor images ($\sim$0.7), while it is weaker with AIA 304 \AA{} ($\sim$0.5), likely because this channel is also sensitive to hotter plasma in the PCTR region. 
In all the $\sim$39\% cases, where CMEs maintain their three-part structure, CME cores could be continuously traced outward from the inner corona into the LASCO/C2 FOV, sometimes as far as ~6 R$_\odot$, strongly reinforcing their prominence origin.

We further back-tracked CME cores from LASCO/C2 FOV to their on-disk source regions under the assumption of no inner coronal observations and by adopting linear kinematics with constant velocity. 
This exercise revealed offsets of up to 40° in position angle and up to 140 minutes in onset time between the actual and back-extrapolated values. 
These results highlight the critical importance of inner coronal observations for reliably connecting CMEs with their source structures.

In summary, using inner-coronal observations of carefully selected prominence-associated limb events, we demonstrate that the CME core in the inner corona is predominantly composed of prominence material, which can be indirectly traced into the outer corona, suggesting its persistence in CME cores at larger heights.

\begin{acknowledgments}
We acknowledge the use of data from the Mauna Loa Solar Observatory (MLSO) K-Cor coronagraph, operated by the High Altitude Observatory as part of the National Center for Atmospheric Research (NCAR). We also thank the Global Oscillation Network Group (GONG), managed by the National Solar Observatory (NSO), for providing the H$\alpha$ observations used in this study. EUV data were obtained from the Solar Dynamics Observatory (SDO)/Atmospheric Imaging Assembly (AIA), a mission of NASA’s Living With a Star (LWS) Program. The SOHO/LASCO observations employed here were produced by an international consortium comprising the Naval Research Laboratory (USA), Max-Planck-Institut für Aeronomie (Germany), Laboratoire d’Astronomie Spatiale (France), and the University of Birmingham (UK); SOHO is a joint project of international cooperation between ESA and NASA. J.J. acknowledges funding support from the SERB-CRG grant (CRG/2023/007464) provided by the Anusandhan National Research Foundation, India.
\end{acknowledgments}


\bibliography{reference}{}

@ARTICLE{Song_2023_1,
       author = {{Song}, Hongqiang and {Li}, Leping and {Zhou}, Zhenjun and {Xia}, Lidong and {Cheng}, Xin and {Chen}, Yao},
        title = "{The Structure of Coronal Mass Ejections Recorded by the K-Coronagraph at Mauna Loa Solar Observatory}",
      journal = {\apjl},
         year = 2023,
        month = jul,
       volume = {952},
       number = {1},
          eid = {L22},
        pages = {L22},
          doi = {10.3847/2041-8213/ace422},
archivePrefix = {arXiv},
       eprint = {2307.01398},
 primaryClass = {astro-ph.SR},
       adsurl = {https://ui.adsabs.harvard.edu/abs/2023ApJ...952L..22S}
}

@ARTICLE{Illing_1985,
       author = {{Illing}, R.~M.~E. and {Hundhausen}, A.~J.},
        title = "{Observation of a coronal transient from 1.2 to 6 solar radii}",
      journal = {\jgr},
         year = 1985,
        month = jan,
       volume = {90},
       number = {A1},
        pages = {275-282},
          doi = {10.1029/JA090iA01p00275},
       adsurl = {https://ui.adsabs.harvard.edu/abs/1985JGR....90..275I}
}

@ARTICLE{Illing_1986,
       author = {{Illing}, R.~M.~E. and {Hundhausen}, A.~J.},
        title = "{Disruption of a coronal streamer by an eruptive prominence and coronal mass ejection}",
      journal = {\jgr},
         year = 1986,
        month = oct,
       volume = {91},
       number = {A10},
        pages = {10951-10960},
          doi = {10.1029/JA091iA10p10951},
       adsurl = {https://ui.adsabs.harvard.edu/abs/1986JGR....9110951I}
}

@ARTICLE{House_1981,
       author = {{House}, L.~L. and {Wagner}, W.~J. and {Hildner}, E. and {Sawyer}, C. and {Schmidt}, H.~U.},
        title = "{Studies of the corona with the Solar Maximum Mission coronagraph/polarimeter}",
      journal = {\apjl},
         year = 1981,
        month = mar,
       volume = {244},
        pages = {L117-L121},
          doi = {10.1086/183494},
       adsurl = {https://ui.adsabs.harvard.edu/abs/1981ApJ...244L.117H}
}

@ARTICLE{Howard_2017,
       author = {{Howard}, T.~A. and {DeForest}, C.~E. and {Schneck}, U.~G. and {Alden}, C.~R.},
        title = "{Challenging Some Contemporary Views of Coronal Mass Ejections. II. The Case for Absent Filaments}",
      journal = {\apj},
         year = 2017,
        month = jan,
       volume = {834},
       number = {1},
          eid = {86},
        pages = {86},
          doi = {10.3847/1538-4357/834/1/86},
       adsurl = {https://ui.adsabs.harvard.edu/abs/2017ApJ...834...86H}
}

@ARTICLE{Song_2019,
       author = {{Song}, H.~Q. and {Zhang}, J. and {Li}, L.~P. and {Liu}, Y.~D. and {Zhu}, B. and {Wang}, B. and {Zheng}, R.~S. and {Chen}, Y.},
        title = "{The Structure of Solar Coronal Mass Ejections in the Extreme-ultraviolet Passbands}",
      journal = {\apj},
         year = 2019,
        month = dec,
       volume = {887},
       number = {2},
          eid = {124},
        pages = {124},
          doi = {10.3847/1538-4357/ab50b6},
archivePrefix = {arXiv},
       eprint = {1910.09735},
 primaryClass = {astro-ph.SR},
       adsurl = {https://ui.adsabs.harvard.edu/abs/2019ApJ...887..124S}
}

@ARTICLE{Subramanian_2001,
       author = {{Subramanian}, Prasad and {Dere}, K.~P.},
        title = "{Source Regions of Coronal Mass Ejections}",
      journal = {\apj},
         year = 2001,
        month = nov,
       volume = {561},
       number = {1},
        pages = {372-395},
          doi = {10.1086/323213},
archivePrefix = {arXiv},
       eprint = {astro-ph/0107138},
 primaryClass = {astro-ph},
       adsurl = {https://ui.adsabs.harvard.edu/abs/2001ApJ...561..372S}
}

@ARTICLE{Satabdwa_2023_1,
       author = {{Majumdar}, Satabdwa and {Patel}, Ritesh and {Pant}, Vaibhav and {Banerjee}, Dipankar and {Rawat}, Aarushi and {Pradhan}, Abhas and {Singh}, Paritosh},
        title = "{A Coronal Mass Ejection Source Region Catalog and Their Associated Properties}",
      journal = {\apjs},
         year = 2023,
        month = sep,
       volume = {268},
       number = {1},
          eid = {38},
        pages = {38},
          doi = {10.3847/1538-4365/aceb62},
archivePrefix = {arXiv},
       eprint = {2307.13208},
 primaryClass = {astro-ph.SR},
       adsurl = {https://ui.adsabs.harvard.edu/abs/2023ApJS..268...38M}
}

@ARTICLE{Poland_1976,
       author = {{Poland}, A.~I. and {Munro}, R.~H.},
        title = "{Interpretation of broad-band polarimetry of solar coronal transients: importance of the Halpha emission.}",
      journal = {\apj},
         year = 1976,
        month = nov,
       volume = {209},
        pages = {927-934},
          doi = {10.1086/154791},
       adsurl = {https://ui.adsabs.harvard.edu/abs/1976ApJ...209..927P}
}

@ARTICLE{Mierla_2011,
       author = {{Mierla}, M. and {Chifu}, I. and {Inhester}, B. and {Rodriguez}, L. and {Zhukov}, A.},
        title = "{Low polarised emission from the core of coronal mass ejections}",
      journal = {\aap},
         year = 2011,
        month = jun,
       volume = {530},
          eid = {L1},
        pages = {L1},
          doi = {10.1051/0004-6361/201016295},
archivePrefix = {arXiv},
       eprint = {1105.3391},
 primaryClass = {astro-ph.SR},
       adsurl = {https://ui.adsabs.harvard.edu/abs/2011A&A...530L...1M}
}

@ARTICLE{Song_2023_2,
       author = {{Song}, Hongqiang and {Zhang}, Jie and {Li}, Leping and {Yang}, Zihao and {Xia}, Lidong and {Zheng}, Ruisheng and {Chen}, Yao},
        title = "{On the Nature of the Three-part Structure of Solar Coronal Mass Ejections}",
      journal = {\apj},
         year = 2023,
        month = jan,
       volume = {942},
       number = {1},
          eid = {19},
        pages = {19},
          doi = {10.3847/1538-4357/aca6e0},
archivePrefix = {arXiv},
       eprint = {2212.04013},
 primaryClass = {astro-ph.SR},
       adsurl = {https://ui.adsabs.harvard.edu/abs/2023ApJ...942...19S}
}

@ARTICLE{Howard_2015,
       author = {{Howard}, T.~A.},
        title = "{Measuring an Eruptive Prominence at Large Distances from the Sun. I. Ionization and Early Evolution}",
      journal = {\apj},
         year = 2015,
        month = jun,
       volume = {806},
       number = {2},
          eid = {175},
        pages = {175},
          doi = {10.1088/0004-637X/806/2/175},
       adsurl = {https://ui.adsabs.harvard.edu/abs/2015ApJ...806..175H}
}

@ARTICLE{Mierla_2024,
       author = {{Mierla}, M. and {Zhukov}, A.~N. and {Berghmans}, D. and {Parenti}, S. and {Auch{\`e}re}, F. and {Heinzel}, P. and {Seaton}, D.~B. and {Palmerio}, E. and {Jej{\v{c}}i{\v{c}}}, S. and {Janssens}, J. and {Kraaikamp}, E. and {Nicula}, B. and {Long}, D.~M. and {Hayes}, L.~A. and {Jebaraj}, I.~C. and {Talpeanu}, D. -C. and {D'Huys}, E. and {Dolla}, L. and {Gissot}, S. and {Magdaleni{\'c}}, J. and {Rodriguez}, L. and {Shestov}, S. and {Stegen}, K. and {Verbeeck}, C. and {Sasso}, C. and {Romoli}, M. and {Andretta}, V.},
        title = "{Prominence eruption observed in He II 304 {\r{A}} up to >6 R$_{{\ensuremath{\odot}}}$ by EUI/FSI aboard Solar Orbiter}",
      journal = {\aap},
         year = 2022,
        month = jun,
       volume = {662},
          eid = {L5},
        pages = {L5},
          doi = {10.1051/0004-6361/202244020},
archivePrefix = {arXiv},
       eprint = {2205.15214},
 primaryClass = {astro-ph.SR},
       adsurl = {https://ui.adsabs.harvard.edu/abs/2022A&A...662L...5M}
}

@ARTICLE{Song_2019_1,
       author = {{Song}, H.~Q. and {Zhang}, J. and {Cheng}, X. and {Li}, L.~P. and {Tang}, Y.~Z. and {Wang}, B. and {Zheng}, R.~S. and {Chen}, Y.},
        title = "{On the Nature of the Bright Core of Solar Coronal Mass Ejections}",
      journal = {\apj},
         year = 2019,
        month = sep,
       volume = {883},
       number = {1},
          eid = {43},
        pages = {43},
          doi = {10.3847/1538-4357/ab304c},
       adsurl = {https://ui.adsabs.harvard.edu/abs/2019ApJ...883...43S}
}

@ARTICLE{Forbes_2000,
       author = {{Forbes}, T.~G.},
        title = "{A review on the genesis of coronal mass ejections}",
      journal = {\jgr},
         year = 2000,
        month = oct,
       volume = {105},
       number = {A10},
        pages = {23153-23166},
          doi = {10.1029/2000JA000005},
       adsurl = {https://ui.adsabs.harvard.edu/abs/2000JGR...10523153F}
}

@ARTICLE{Fuller_2008,
       author = {{Fuller}, J. and {Gibson}, S.~E. and {de Toma}, G. and {Fan}, Y.},
        title = "{Observing the Unobservable? Modeling Coronal Cavity Densities}",
      journal = {\apj},
         year = 2008,
        month = may,
       volume = {678},
       number = {1},
        pages = {515-530},
          doi = {10.1086/533527},
       adsurl = {https://ui.adsabs.harvard.edu/abs/2008ApJ...678..515F}
}

@ARTICLE{howard_deforest_2012,
       author = {{Howard}, T.~A. and {DeForest}, C.~E.},
        title = "{Inner Heliospheric Flux Rope Evolution via Imaging of Coronal Mass Ejections}",
      journal = {\apj},
         year = 2012,
        month = feb,
       volume = {746},
       number = {1},
          eid = {64},
        pages = {64},
          doi = {10.1088/0004-637X/746/1/64},
       adsurl = {https://ui.adsabs.harvard.edu/abs/2012ApJ...746...64H}
}

@ARTICLE{Vourildas_2013,
       author = {{Vourlidas}, A. and {Lynch}, B.~J. and {Howard}, R.~A. and {Li}, Y.},
        title = "{How Many CMEs Have Flux Ropes? Deciphering the Signatures of Shocks, Flux Ropes, and Prominences in Coronagraph Observations of CMEs}",
      journal = {\solphys},
         year = 2013,
        month = may,
       volume = {284},
       number = {1},
        pages = {179-201},
          doi = {10.1007/s11207-012-0084-8},
archivePrefix = {arXiv},
       eprint = {1207.1599},
 primaryClass = {astro-ph.SR},
       adsurl = {https://ui.adsabs.harvard.edu/abs/2013SoPh..284..179V}
}

@ARTICLE{Ritesh_2022,
       author = {{Patel}, Ritesh and {Majumdar}, Satabdwa and {Pant}, Vaibhav and {Banerjee}, Dipankar},
        title = "{A Simple Radial Gradient Filter for Batch-Processing of Coronagraph Images}",
      journal = {\solphys},
         year = 2022,
        month = mar,
       volume = {297},
       number = {3},
          eid = {27},
        pages = {27},
          doi = {10.1007/s11207-022-01957-y},
archivePrefix = {arXiv},
       eprint = {2201.13043},
 primaryClass = {astro-ph.SR},
       adsurl = {https://ui.adsabs.harvard.edu/abs/2022SoPh..297...27P}
}

@ARTICLE{Munro_1979,
       author = {{Munro}, R.~H. and {Gosling}, J.~T. and {Hildner}, E. and {MacQueen}, R.~M. and {Poland}, A.~I. and {Ross}, C.~L.},
        title = "{The association of coronal mass ejection transients with other forms of solar activity.}",
      journal = {\solphys},
         year = 1979,
        month = feb,
       volume = {61},
       number = {1},
        pages = {201-215},
          doi = {10.1007/BF00155456},
       adsurl = {https://ui.adsabs.harvard.edu/abs/1979SoPh...61..201M}
}

@ARTICLE{Hansen_1971,
       author = {{Hansen}, R.~T. and {Garcia}, C.~J. and {Grognard}, R.~J. -M. and {Sheridan}, K.~V.},
        title = "{A coronal disturbance observed simultaneously with a white-light corona-meter and the 80 MHz Culgoora radioheliograph}",
      journal = {\pasa},
         year = 1971,
        month = jul,
       volume = {2},
        pages = {57},
          doi = {10.1017/S1323358000012856},
       adsurl = {https://ui.adsabs.harvard.edu/abs/1971PASA....2...57H}
}

@article{Webb2012,
  author = {David F. Webb and Timothy A. Howard},
  title = {Coronal Mass Ejections: Observations},
  journal = {Living Reviews in Solar Physics},
  year = {2012},
  volume = {9},
  number = {1},
  pages = {3},
  doi = {10.12942/lrsp-2012-3},
  url = {https://doi.org/10.12942/lrsp-2012-3}
}

@ARTICLE{Schmahl_1977,
       author = {{Schmahl}, E. and {Hildner}, E.},
        title = "{Coronal mass-ejections-kinematics of the 19 December 1973 event.}",
      journal = {\solphys},
         year = 1977,
        month = dec,
       volume = {55},
       number = {2},
        pages = {473-490},
          doi = {10.1007/BF00152588},
       adsurl = {https://ui.adsabs.harvard.edu/abs/1977SoPh...55..473S}
}

@ARTICLE{Burlaga_1998,
       author = {{Burlaga}, L. and {Fitzenreiter}, R. and {Lepping}, R. and {Ogilvie}, K. and {Szabo}, A. and {Lazarus}, A. and {Steinberg}, J. and {Gloeckler}, G. and {Howard}, R. and {Michels}, D. and {Farrugia}, C. and {Lin}, R.~P. and {Larson}, D.~E.},
        title = "{A magnetic cloud containing prominence material: January 1997}",
      journal = {\jgr},
         year = 1998,
        month = jan,
       volume = {103},
       number = {A1},
        pages = {277-286},
          doi = {10.1029/97JA02768},
       adsurl = {https://ui.adsabs.harvard.edu/abs/1998JGR...103..277B}
}

@ARTICLE{Skoug_1999,
       author = {{Skoug}, R.~M. and {Bame}, S.~J. and {Feldman}, W.~C. and {Gosling}, J.~T. and {McComas}, D.~J. and {Steinberg}, J.~T. and {Tokar}, R.~L. and {Riley}, P. and {Burlaga}, L.~F. and {Ness}, N.~F. and {Smith}, C.~W.},
        title = "{A prolonged He$^{+}$ enhancement within a coronal mass ejection in the solar wind}",
      journal = {\grl},
         year = 1999,
        month = jan,
       volume = {26},
       number = {2},
        pages = {161-164},
          doi = {10.1029/1998GL900207},
       adsurl = {https://ui.adsabs.harvard.edu/abs/1999GeoRL..26..161S}
}

@ARTICLE{Gopalswwamy_1998,
       author = {{Gopalswamy}, N. and {Hanaoka}, Y.},
        title = "{Coronal Dimming Associated with a Giant Prominence Eruption}",
      journal = {\apjl},
         year = 1998,
        month = may,
       volume = {498},
       number = {2},
        pages = {L179-L182},
          doi = {10.1086/311330},
       adsurl = {https://ui.adsabs.harvard.edu/abs/1998ApJ...498L.179G}
}

@ARTICLE{Gilbert_2000,
       author = {{Gilbert}, Holly R. and {Holzer}, Thomas E. and {Burkepile}, Joan T. and {Hundhausen}, Arthur J.},
        title = "{Active and Eruptive Prominences and Their Relationship to Coronal Mass Ejections}",
      journal = {\apj},
         year = 2000,
        month = jul,
       volume = {537},
       number = {1},
        pages = {503-515},
          doi = {10.1086/309030},
       adsurl = {https://ui.adsabs.harvard.edu/abs/2000ApJ...537..503G}
}

@ARTICLE{Gopalswamy_2003,
       author = {{Gopalswamy}, N. and {Shimojo}, M. and {Lu}, W. and {Yashiro}, S. and {Shibasaki}, K. and {Howard}, R.~A.},
        title = "{Prominence Eruptions and Coronal Mass Ejection: A Statistical Study Using Microwave Observations}",
      journal = {\apj},
         year = 2003,
        month = mar,
       volume = {586},
       number = {1},
        pages = {562-578},
          doi = {10.1086/367614},
       adsurl = {https://ui.adsabs.harvard.edu/abs/2003ApJ...586..562G}
}

@ARTICLE{Flippov_2008,
       author = {{Filippov}, B. and {Koutchmy}, S.},
        title = "{Causal relationships between eruptive prominences and coronal mass ejections}",
      journal = {Annales Geophysicae},
         year = 2008,
        month = oct,
       volume = {26},
       number = {10},
        pages = {3025-3031},
          doi = {10.5194/angeo-26-3025-2008},
archivePrefix = {arXiv},
       eprint = {0711.4752},
 primaryClass = {astro-ph},
       adsurl = {https://ui.adsabs.harvard.edu/abs/2008AnGeo..26.3025F}
}

@article{Webb1987,
  author = {D. F. Webb and A. J. Hundhausen},
  title = {Activity associated with the solar origin of coronal mass ejections},
  journal = {Solar Physics},
  year = {1987},
  volume = {108},
  number = {2},
  pages = {383--401},
  doi = {10.1007/BF00214170},
  url = {https://doi.org/10.1007/BF00214170}
}

@article{Maricic2009,
  author = {Darije Maričić and Bojan Vršnak and Dragan Roša},
  title = {Relative Kinematics of the Leading Edge and the Prominence in Coronal Mass Ejections},
  journal = {Solar Physics},
  year = {2009},
  volume = {260},
  number = {1},
  pages = {177--189},
  doi = {10.1007/s11207-009-9421-y},
  url = {https://doi.org/10.1007/s11207-009-9421-y}
}

@article{Maricic2004,
  author = {D. Maričić and B. Vršnak and A. L. Stanger and A. Veronig},
  title = {Coronal Mass Ejection of 15 May 2001: I. Evolution of Morphological Features of the Eruption},
  journal = {Solar Physics},
  year = {2004},
  volume = {225},
  number = {2},
  pages = {337--353},
  doi = {10.1007/s11207-004-3748-1},
  url = {https://doi.org/10.1007/s11207-004-3748-1}
}

@ARTICLE{Hori_2002,
       author = {{Hori}, K. and {Culhane}, J.~L.},
        title = "{Trajectories of microwave prominence eruptions}",
      journal = {\aap},
         year = 2002,
        month = feb,
       volume = {382},
        pages = {666-677},
          doi = {10.1051/0004-6361:20011658},
       adsurl = {https://ui.adsabs.harvard.edu/abs/2002A&A...382..666H}
}

@INPROCEEDINGS{wang_1998,
       author = {{Wang}, Haimin and {Goode}, Philip R.},
        title = "{Synoptic Observing Programs at Big Bear Solar Observatory}",
    booktitle = {Synoptic Solar Physics},
         year = 1998,
       editor = {{Balasubramaniam}, K.~S. and {Harvey}, Jack and {Rabin}, D.},
       series = {Astronomical Society of the Pacific Conference Series},
       volume = {140},
        month = jan,
        pages = {497},
       adsurl = {https://ui.adsabs.harvard.edu/abs/1998ASPC..140..497W}
}

@ARTICLE{Song_2014,
       author = {{Song}, H.~Q. and {Zhang}, J. and {Chen}, Y. and {Cheng}, X.},
        title = "{Direct Observations of Magnetic Flux Rope Formation during a Solar Coronal Mass Ejection}",
      journal = {\apjl},
         year = 2014,
        month = sep,
       volume = {792},
       number = {2},
          eid = {L40},
        pages = {L40},
          doi = {10.1088/2041-8205/792/2/L40},
archivePrefix = {arXiv},
       eprint = {1408.2000},
 primaryClass = {astro-ph.SR},
       adsurl = {https://ui.adsabs.harvard.edu/abs/2014ApJ...792L..40S}
}

@ARTICLE{Lepri_2010,
       author = {{Lepri}, S.~T. and {Zurbuchen}, T.~H.},
        title = "{Direct Observational Evidence of Filament Material Within Interplanetary Coronal Mass Ejections}",
      journal = {\apjl},
         year = 2010,
        month = nov,
       volume = {723},
       number = {1},
        pages = {L22-L27},
          doi = {10.1088/2041-8205/723/1/L22},
       adsurl = {https://ui.adsabs.harvard.edu/abs/2010ApJ...723L..22L}
}

@ARTICLE{Wood_2016,
       author = {{Wood}, Brian E. and {Howard}, Russell A. and {Linton}, Mark G.},
        title = "{Imaging Prominence Eruptions out to 1 AU}",
      journal = {\apj},
         year = 2016,
        month = jan,
       volume = {816},
       number = {2},
          eid = {67},
        pages = {67},
          doi = {10.3847/0004-637X/816/2/67},
archivePrefix = {arXiv},
       eprint = {1512.06748},
 primaryClass = {astro-ph.SR},
       adsurl = {https://ui.adsabs.harvard.edu/abs/2016ApJ...816...67W}
}

@ARTICLE{kcor_2017,
       author = {{Thompson}, W.~T. and {St. Cyr}, O.~C. and {Burkepile}, J.~T. and {Posner}, A.},
        title = "{Automatic Near-Real-Time Detection of CMEs in Mauna Loa K-Cor Coronagraph Images}",
      journal = {Space Weather},
         year = 2017,
        month = oct,
       volume = {15},
       number = {10},
        pages = {1288-1299},
          doi = {10.1002/2017SW001694},
       adsurl = {https://ui.adsabs.harvard.edu/abs/2017SpWea..15.1288T}
}

@article{AIA_2012,
  author = {James R. Lemen and Alan M. Title and David J. Akin and Paul F. Boerner and Catherine Chou and Jerry F. Drake and Dexter W. Duncan and Christopher G. Edwards and Frank M. Friedlaender and Gary F. Heyman and Neal E. Hurlburt and Noah L. Katz and Gary D. Kushner and Michael Levay and Russell W. Lindgren and Dnyanesh P. Mathur and Edward L. McFeaters and Sarah Mitchell and Roger A. Rehse and Carolus J. Schrijver and Larry A. Springer and Robert A. Stern and Theodore D. Tarbell and Jean-Pierre Wuelser and C. Jacob Wolfson and Carl Yanari and Jay A. Bookbinder and Peter N. Cheimets and David Caldwell and Edward E. Deluca and Richard Gates and Leon Golub and Sang Park and William A. Podgorski and Rock I. Bush and Philip H. Scherrer and Mark A. Gummin and Peter Smith and Gary Auker and Paul Jerram and Peter Pool and Regina Soufli and David L. Windt and Sarah Beardsley and Matthew Clapp and James Lang and Nicholas Waltham},
  title = {The Atmospheric Imaging Assembly (AIA) on the Solar Dynamics Observatory (SDO)},
  journal = {Solar Physics},
  year = {2012},
  volume = {275},
  number = {1},
  pages = {17--40},
  doi = {10.1007/s11207-011-9776-8},
  url = {https://doi.org/10.1007/s11207-011-9776-8}
}

@ARTICLE{Gibson_2006,
       author = {{Gibson}, S.~E. and {Foster}, D. and {Burkepile}, J. and {de Toma}, G. and {Stanger}, A.},
        title = "{The Calm before the Storm: The Link between Quiescent Cavities and Coronal Mass Ejections}",
      journal = {\apj},
         year = 2006,
        month = apr,
       volume = {641},
       number = {1},
        pages = {590-605},
          doi = {10.1086/500446},
       adsurl = {https://ui.adsabs.harvard.edu/abs/2006ApJ...641..590G}
}

@article{lyot1930,
  title={La couronne solaire {\'e}tudi{\'e}e en dehors des {\'e}clipses},
  author={Lyot, Bernard},
  journal={Comptes rendus},
  volume={191},
  pages={834--837},
  year={1930}
}

@ARTICLE{chen_2011,
       author = {{Chen}, P.~F.},
        title = "{Coronal Mass Ejections: Models and Their Observational Basis}",
      journal = {Living Reviews in Solar Physics},
         year = 2011,
        month = dec,
       volume = {8},
       number = {1},
          eid = {1},
        pages = {1},
          doi = {10.12942/lrsp-2011-1},
       adsurl = {https://ui.adsabs.harvard.edu/abs/2011LRSP....8....1C}
}

@article{Gopalswamy_2009,
  author    = {Gopalswamy, N. and Yashiro, S. and Michalek, G. and Stenborg, G. and Vourlidas, A. and Freeland, S. and Howard, R.},
  title     = {The SOHO/LASCO CME Catalog},
  journal   = {Earth, Moon, and Planets},
  year      = {2009},
  volume    = {104},
  number    = {1},
  pages     = {295--313},
  doi       = {10.1007/s11038-008-9282-7},
  url       = {https://doi.org/10.1007/s11038-008-9282-7},
  issn      = {1573-0794}
}

@ARTICLE{Temmer_2021,
       author = {{Temmer}, Manuela},
        title = "{Space weather: the solar perspective: An update to Schwenn (2006)}",
      journal = {Living Reviews in Solar Physics},
         year = 2021,
        month = dec,
       volume = {18},
       number = {1},
          eid = {4},
        pages = {4},
          doi = {10.1007/s41116-021-00030-3},
archivePrefix = {arXiv},
       eprint = {2104.04261},
 primaryClass = {astro-ph.SR},
       adsurl = {https://ui.adsabs.harvard.edu/abs/2021LRSP...18....4T}
}

@ARTICLE{Schwenn_2006,
       author = {{Schwenn}, Rainer},
        title = "{Space Weather: The Solar Perspective}",
      journal = {Living Reviews in Solar Physics},
         year = 2006,
        month = dec,
       volume = {3},
       number = {1},
          eid = {2},
        pages = {2},
          doi = {10.12942/lrsp-2006-2},
       adsurl = {https://ui.adsabs.harvard.edu/abs/2006LRSP....3....2S}
}

@ARTICLE{KCORpaepr2016,
       author = {{Tomczyk}, S. and {Landi}, E. and {Burkepile}, J.~T. and {Casini}, R. and {DeLuca}, E.~E. and {Fan}, Y. and {Gibson}, S.~E. and {Lin}, H. and {McIntosh}, S.~W. and {Solomon}, S.~C. and {de Toma}, G. and {de Wijn}, A.~G. and {Zhang}, J.},
        title = "{Scientific objectives and capabilities of the Coronal Solar Magnetism Observatory}",
      journal = {Journal of Geophysical Research (Space Physics)},
         year = 2016,
        month = aug,
       volume = {121},
       number = {8},
        pages = {7470-7487},
          doi = {10.1002/2016JA022871},
       adsurl = {https://ui.adsabs.harvard.edu/abs/2016JGRA..121.7470T}
}

@INPROCEEDINGS{stereo_euvi_2004,
       author = {{Wuelser}, Jean-Pierre and {Lemen}, James R. and {Tarbell}, Theodore D. and {Wolfson}, C.~J. and {Cannon}, Joseph C. and {Carpenter}, Brock A. and {Duncan}, Dexter W. and {Gradwohl}, Glenn S. and {Meyer}, Syndie B. and {Moore}, Augustus S. and {Navarro}, Rosemarie L. and {Pearson}, J.~D. and {Rossi}, George R. and {Springer}, Larry A. and {Howard}, Russell A. and {Moses}, John D. and {Newmark}, Jeffrey S. and {Delaboudiniere}, Jean-Pierre and {Artzner}, Guy E. and {Auchere}, Frederic and {Bougnet}, Marie and {Bouyries}, Philippe and {Bridou}, Francoise and {Clotaire}, Jean-Yves and {Colas}, Gerard and {Delmotte}, Franck and {Jerome}, Arnaud and {Lamare}, Michel and {Mercier}, Raymond and {Mullot}, Michel and {Ravet}, Marie-Francoise and {Song}, Xueyan and {Bothmer}, Volker and {Deutsch}, Werner},
        title = "{EUVI: the STEREO-SECCHI extreme ultraviolet imager}",
    booktitle = {Telescopes and Instrumentation for Solar Astrophysics},
         year = 2004,
       editor = {{Fineschi}, Silvano and {Gummin}, Mark A.},
       series = {Society of Photo-Optical Instrumentation Engineers (SPIE) Conference Series},
       volume = {5171},
        month = feb,
        pages = {111-122},
          doi = {10.1117/12.506877},
       adsurl = {https://ui.adsabs.harvard.edu/abs/2004SPIE.5171..111W}
}

@INPROCEEDINGS{GONG_2011,
       author = {{Harvey}, J.~W. and {Bolding}, J. and {Clark}, R. and {Hauth}, D. and {Hill}, F. and {Kroll}, R. and {Luis}, G. and {Mills}, N. and {Purdy}, T. and {Henney}, C. and {Holland}, D. and {Winter}, J.},
        title = "{Full-disk Solar H-alpha Images From GONG}",
    booktitle = {AAS/Solar Physics Division Abstracts \#42},
         year = 2011,
       series = {AAS/Solar Physics Division Meeting},
       volume = {42},
        month = may,
          eid = {17.45},
        pages = {17.45},
       adsurl = {https://ui.adsabs.harvard.edu/abs/2011SPD....42.1745H}
}

@ARTICLE{LASCO_1995,
       author = {{Brueckner}, G.~E. and {Howard}, R.~A. and {Koomen}, M.~J. and {Korendyke}, C.~M. and {Michels}, D.~J. and {Moses}, J.~D. and {Socker}, D.~G. and {Dere}, K.~P. and {Lamy}, P.~L. and {Llebaria}, A. and {Bout}, M.~V. and {Schwenn}, R. and {Simnett}, G.~M. and {Bedford}, D.~K. and {Eyles}, C.~J.},
        title = "{The Large Angle Spectroscopic Coronagraph (LASCO)}",
      journal = {\solphys},
         year = 1995,
        month = dec,
       volume = {162},
       number = {1-2},
        pages = {357-402},
          doi = {10.1007/BF00733434},
       adsurl = {https://ui.adsabs.harvard.edu/abs/1995SoPh..162..357B}
}

@ARTICLE{SOHO_1995,
       author = {{Domingo}, V. and {Fleck}, B. and {Poland}, A.~I.},
        title = "{The SOHO Mission: an Overview}",
      journal = {\solphys},
         year = 1995,
        month = dec,
       volume = {162},
       number = {1-2},
        pages = {1-37},
          doi = {10.1007/BF00733425},
       adsurl = {https://ui.adsabs.harvard.edu/abs/1995SoPh..162....1D}
}

@ARTICLE{Zhang_2012,
       author = {{Zhang}, Jie and {Cheng}, Xin and {Ding}, Ming-De},
        title = "{Observation of an evolving magnetic flux rope before and during a solar eruption}",
      journal = {Nature Communications},
         year = 2012,
        month = mar,
       volume = {3},
          eid = {747},
        pages = {747},
          doi = {10.1038/ncomms1753},
archivePrefix = {arXiv},
       eprint = {1203.4859},
 primaryClass = {astro-ph.SR},
       adsurl = {https://ui.adsabs.harvard.edu/abs/2012NatCo...3..747Z}
}

@ARTICLE{Peat_2024,
       author = {{Peat}, Aaron W. and {Osborne}, Christopher M.~J. and {Heinzel}, Petr},
        title = "{Doppler dimming and brightening effects in solar prominences}",
      journal = {\mnras},
         year = 2024,
        month = sep,
       volume = {533},
       number = {1},
        pages = {L19-L24},
          doi = {10.1093/mnrasl/slae055},
archivePrefix = {arXiv},
       eprint = {2406.12551},
 primaryClass = {astro-ph.SR},
       adsurl = {https://ui.adsabs.harvard.edu/abs/2024MNRAS.533L..19P}
}

@ARTICLE{boerner_2012,
       author = {{Boerner}, Paul and {Edwards}, Christopher and {Lemen}, James and {Rausch}, Adam and {Schrijver}, Carolus and {Shine}, Richard and {Shing}, Lawrence and {Stern}, Robert and {Tarbell}, Theodore and {Title}, Alan and {Wolfson}, C. Jacob and {Soufli}, Regina and {Spiller}, Eberhard and {Gullikson}, Eric and {McKenzie}, David and {Windt}, David and {Golub}, Leon and {Podgorski}, William and {Testa}, Paola and {Weber}, Mark},
        title = "{Initial Calibration of the Atmospheric Imaging Assembly (AIA) on the Solar Dynamics Observatory (SDO)}",
      journal = {\solphys},
         year = 2012,
        month = jan,
       volume = {275},
       number = {1-2},
        pages = {41-66},
          doi = {10.1007/s11207-011-9804-8},
       adsurl = {https://ui.adsabs.harvard.edu/abs/2012SoPh..275...41B}
}

@ARTICLE{Rivera_2019,
       author = {{Rivera}, Yeimy J. and {Landi}, Enrico and {Lepri}, Susan T.},
        title = "{Identifying Spectral Lines to Study Coronal Mass Ejection Evolution in the Lower Corona}",
      journal = {\apjs},
         year = 2019,
        month = aug,
       volume = {243},
       number = {2},
          eid = {34},
        pages = {34},
          doi = {10.3847/1538-4365/ab2bfe},
       adsurl = {https://ui.adsabs.harvard.edu/abs/2019ApJS..243...34R}
}

@ARTICLE{Vial_2012,
       author = {{Vial}, J. -C. and {Olivier}, K. and {Philippon}, A.~A. and {Vourlidas}, A. and {Yurchyshyn}, V.},
        title = "{High spatial resolution VAULT H-Ly{\ensuremath{\alpha}} observations and multiwavelength analysis of an active region filament}",
      journal = {\aap},
         year = 2012,
        month = may,
       volume = {541},
          eid = {A108},
        pages = {A108},
          doi = {10.1051/0004-6361/201118275},
       adsurl = {https://ui.adsabs.harvard.edu/abs/2012A&A...541A.108V}
}

@ARTICLE{Van_1950,
       author = {{van de Hulst}, H.~C.},
        title = "{The electron density of the solar corona}",
      journal = {\bain},
         year = 1950,
        month = feb,
       volume = {11},
        pages = {135},
       adsurl = {https://ui.adsabs.harvard.edu/abs/1950BAN....11..135V}
}

@BOOK{Billing_1966,
       author = {{Billings}, Donald E.},
        title = "{A guide to the solar corona}",
         year = 1966,
       adsurl = {https://ui.adsabs.harvard.edu/abs/1966gtsc.book.....B}
}

@ARTICLE{Morgan_2006,
       author = {{Morgan}, Huw and {Habbal}, Shadia Rifai and {Woo}, Richard},
        title = "{The Depiction of Coronal Structure in White-Light Images}",
      journal = {\solphys},
         year = 2006,
        month = jul,
       volume = {236},
       number = {2},
        pages = {263-272},
          doi = {10.1007/s11207-006-0113-6},
archivePrefix = {arXiv},
       eprint = {astro-ph/0602174},
 primaryClass = {astro-ph},
       adsurl = {https://ui.adsabs.harvard.edu/abs/2006SoPh..236..263M}
}

@ARTICLE{Yashiro_2004,
       author = {{Yashiro}, S. and {Gopalswamy}, N. and {Michalek}, G. and {St. Cyr}, O.~C. and {Plunkett}, S.~P. and {Rich}, N.~B. and {Howard}, R.~A.},
        title = "{A catalog of white light coronal mass ejections observed by the SOHO spacecraft}",
      journal = {Journal of Geophysical Research (Space Physics)},
         year = 2004,
        month = jul,
       volume = {109},
       number = {A7},
          eid = {A07105},
        pages = {A07105},
          doi = {10.1029/2003JA010282},
       adsurl = {https://ui.adsabs.harvard.edu/abs/2004JGRA..109.7105Y}
}

@ARTICLE{wang_2011,
       author = {{Wang}, Yuming and {Chen}, Caixia and {Gui}, Bin and {Shen}, Chenglong and {Ye}, Pinzhong and {Wang}, S.},
        title = "{Statistical study of coronal mass ejection source locations: Understanding CMEs viewed in coronagraphs}",
      journal = {Journal of Geophysical Research (Space Physics)},
         year = 2011,
        month = apr,
       volume = {116},
       number = {A4},
          eid = {A04104},
        pages = {A04104},
          doi = {10.1029/2010JA016101},
archivePrefix = {arXiv},
       eprint = {1101.0641},
 primaryClass = {astro-ph.SR},
       adsurl = {https://ui.adsabs.harvard.edu/abs/2011JGRA..116.4104W}
}

@ARTICLE{Skylab_1974,
       author = {{Gosling}, J.~T. and {Hildner}, E. and {MacQueen}, R.~M. and {Munro}, R.~H. and {Poland}, A.~I. and {Ross}, C.~L.},
        title = "{Mass ejections from the Sun: A view from Skylab}",
      journal = {\jgr},
         year = 1974,
        month = nov,
       volume = {79},
       number = {31},
        pages = {4581},
          doi = {10.1029/JA079i031p04581},
       adsurl = {https://ui.adsabs.harvard.edu/abs/1974JGR....79.4581G}
}

@ARTICLE{Bien_2011,
       author = {{Bein}, B.~M. and {Berkebile-Stoiser}, S. and {Veronig}, A.~M. and {Temmer}, M. and {Muhr}, N. and {Kienreich}, I. and {Utz}, D. and {Vr{\v{s}}nak}, B.},
        title = "{Impulsive Acceleration of Coronal Mass Ejections. I. Statistics and Coronal Mass Ejection Source Region Characteristics}",
      journal = {\apj},
         year = 2011,
        month = sep,
       volume = {738},
       number = {2},
          eid = {191},
        pages = {191},
          doi = {10.1088/0004-637X/738/2/191},
archivePrefix = {arXiv},
       eprint = {1108.0561},
 primaryClass = {astro-ph.SR},
       adsurl = {https://ui.adsabs.harvard.edu/abs/2011ApJ...738..191B}
}

@INPROCEEDINGS{cor1_2003,
       author = {{Thompson}, William T. and {Davila}, Joseph M. and {Fisher}, Richard R. and {Orwig}, Larry E. and {Mentzell}, John E. and {Hetherington}, Samuel E. and {Derro}, Rebecca J. and {Federline}, Robert E. and {Clark}, David C. and {Chen}, Philip T.~C. and {Tveekrem}, June L. and {Martino}, Anthony J. and {Novello}, Joseph and {Wesenberg}, Richard P. and {StCyr}, Orville C. and {Reginald}, Nelson L. and {Howard}, Russell A. and {Mehalick}, Kimberly I. and {Hersh}, Michael J. and {Newman}, Miles D. and {Thomas}, Debbie L. and {Card}, Gregory L. and {Elmore}, David F.},
        title = "{COR1 inner coronagraph for STEREO-SECCHI}",
    booktitle = {Innovative Telescopes and Instrumentation for Solar Astrophysics},
         year = 2003,
       editor = {{Keil}, Stephen L. and {Avakyan}, Sergey V.},
       series = {Society of Photo-Optical Instrumentation Engineers (SPIE) Conference Series},
       volume = {4853},
        month = feb,
        pages = {1-11},
          doi = {10.1117/12.460267},
       adsurl = {https://ui.adsabs.harvard.edu/abs/2003SPIE.4853....1T}
}

@ARTICLE{Labrosse_2010,
       author = {{Labrosse}, N. and {Heinzel}, P. and {Vial}, J. -C. and {Kucera}, T. and {Parenti}, S. and {Gun{\'a}r}, S. and {Schmieder}, B. and {Kilper}, G.},
        title = "{Physics of Solar Prominences: I{\textemdash}Spectral Diagnostics and Non-LTE Modelling}",
      journal = {\ssr},
         year = 2010,
        month = apr,
       volume = {151},
       number = {4},
        pages = {243-332},
          doi = {10.1007/s11214-010-9630-6},
archivePrefix = {arXiv},
       eprint = {1001.1620},
 primaryClass = {astro-ph.SR},
       adsurl = {https://ui.adsabs.harvard.edu/abs/2010SSRv..151..243L}
}

@ARTICLE{song_2022,
       author = {{Song}, Hongqiang and {Li}, Leping and {Chen}, Yao},
        title = "{Toward a Unified Explanation for the Three-part Structure of Solar Coronal Mass Ejections}",
      journal = {\apj},
         year = 2022,
        month = jul,
       volume = {933},
       number = {1},
          eid = {68},
        pages = {68},
          doi = {10.3847/1538-4357/ac7239},
archivePrefix = {arXiv},
       eprint = {2205.11682},
 primaryClass = {astro-ph.SR},
       adsurl = {https://ui.adsabs.harvard.edu/abs/2022ApJ...933...68S}
}

@ARTICLE{Illing&athay_1986,
       author = {{Illing}, R.~M.~E. and {Athay}, G.},
        title = "{Physical Conditions in Eruptive Prominences at Several Solar Radii}",
      journal = {\solphys},
         year = 1986,
        month = may,
       volume = {105},
       number = {1},
        pages = {173-190},
          doi = {10.1007/BF00156385},
       adsurl = {https://ui.adsabs.harvard.edu/abs/1986SoPh..105..173I}
}

@ARTICLE{Pooja_2025,
       author = {{Devi}, Pooja and {Gopalswamy}, Nat and {Yashiro}, Seiji and {Akiyama}, Sachiko and {Chandra}, Ramesh and {Koleva}, Kostadinka},
        title = "{Relationship between prominence eruptions and coronal mass ejections during solar cycle 24}",
      journal = {Journal of Astrophysics and Astronomy},
         year = 2025,
        month = aug,
       volume = {46},
       number = {2},
          eid = {64},
        pages = {64},
          doi = {10.1007/s12036-025-10088-2},
archivePrefix = {arXiv},
       eprint = {2505.24202},
 primaryClass = {astro-ph.SR},
       adsurl = {https://ui.adsabs.harvard.edu/abs/2025JApA...46...64D}
}

@ARTICLE{Comp_2008,
       author = {{Tomczyk}, S. and {Card}, G.~L. and {Darnell}, T. and {Elmore}, D.~F. and {Lull}, R. and {Nelson}, P.~G. and {Streander}, K.~V. and {Burkepile}, J. and {Casini}, R. and {Judge}, P.~G.},
        title = "{An Instrument to Measure Coronal Emission Line Polarization}",
      journal = {\solphys},
         year = 2008,
        month = feb,
       volume = {247},
       number = {2},
        pages = {411-428},
          doi = {10.1007/s11207-007-9103-6},
       adsurl = {https://ui.adsabs.harvard.edu/abs/2008SoPh..247..411T}
}

@ARTICLE{Ucomp_2016,
       author = {{Landi}, E. and {Habbal}, S.~R. and {Tomczyk}, S.},
        title = "{Coronal plasma diagnostics from ground-based observations}",
      journal = {Journal of Geophysical Research (Space Physics)},
         year = 2016,
        month = sep,
       volume = {121},
       number = {9},
        pages = {8237-8249},
          doi = {10.1002/2016JA022598},
       adsurl = {https://ui.adsabs.harvard.edu/abs/2016JGRA..121.8237L}
}

@ARTICLE{Solo_eui_2020,
       author = {{Rochus}, P. and {Auch{\`e}re}, F. and {Berghmans}, D. and {Harra}, L. and {Schmutz}, W. and {Sch{\"u}hle}, U. and {Addison}, P. and {Appourchaux}, T. and {Aznar Cuadrado}, R. and {Baker}, D. and {Barbay}, J. and {Bates}, D. and {BenMoussa}, A. and {Bergmann}, M. and {Beurthe}, C. and {Borgo}, B. and {Bonte}, K. and {Bouzit}, M. and {Bradley}, L. and {B{\"u}chel}, V. and {Buchlin}, E. and {B{\"u}chner}, J. and {Cab{\'e}}, F. and {Cadiergues}, L. and {Chaigneau}, M. and {Chares}, B. and {Choque Cortez}, C. and {Coker}, P. and {Condamin}, M. and {Coumar}, S. and {Curdt}, W. and {Cutler}, J. and {Davies}, D. and {Davison}, G. and {Defise}, J. -M. and {Del Zanna}, G. and {Delmotte}, F. and {Delouille}, V. and {Dolla}, L. and {Dumesnil}, C. and {D{\"u}rig}, F. and {Enge}, R. and {Fran{\c{c}}ois}, S. and {Fourmond}, J. -J. and {Gillis}, J. -M. and {Giordanengo}, B. and {Gissot}, S. and {Green}, L.~M. and {Guerreiro}, N. and {Guilbaud}, A. and {Gyo}, M. and {Haberreiter}, M. and {Hafiz}, A. and {Hailey}, M. and {Halain}, J. -P. and {Hansotte}, J. and {Hecquet}, C. and {Heerlein}, K. and {Hellin}, M. -L. and {Hemsley}, S. and {Hermans}, A. and {Hervier}, V. and {Hochedez}, J. -F. and {Houbrechts}, Y. and {Ihsan}, K. and {Jacques}, L. and {J{\'e}r{\^o}me}, A. and {Jones}, J. and {Kahle}, M. and {Kennedy}, T. and {Klaproth}, M. and {Kolleck}, M. and {Koller}, S. and {Kotsialos}, E. and {Kraaikamp}, E. and {Langer}, P. and {Lawrenson}, A. and {Le Clech'}, J. -C. and {Lenaerts}, C. and {Liebecq}, S. and {Linder}, D. and {Long}, D.~M. and {Mampaey}, B. and {Markiewicz-Innes}, D. and {Marquet}, B. and {Marsch}, E. and {Matthews}, S. and {Mazy}, E. and {Mazzoli}, A. and {Meining}, S. and {Meltchakov}, E. and {Mercier}, R. and {Meyer}, S. and {Monecke}, M. and {Monfort}, F. and {Morinaud}, G. and {Moron}, F. and {Mountney}, L. and {M{\"u}ller}, R. and {Nicula}, B. and {Parenti}, S. and {Peter}, H. and {Pfiffner}, D. and {Philippon}, A. and {Phillips}, I. and {Plesseria}, J. -Y. and {Pylyser}, E. and {Rabecki}, F. and {Ravet-Krill}, M. -F. and {Rebellato}, J. and {Renotte}, E. and {Rodriguez}, L. and {Roose}, S. and {Rosin}, J. and {Rossi}, L. and {Roth}, P. and {Rouesnel}, F. and {Roulliay}, M. and {Rousseau}, A. and {Ruane}, K. and {Scanlan}, J. and {Schlatter}, P. and {Seaton}, D.~B. and {Silliman}, K. and {Smit}, S. and {Smith}, P.~J. and {Solanki}, S.~K. and {Spescha}, M. and {Spencer}, A. and {Stegen}, K. and {Stockman}, Y. and {Szwec}, N. and {Tamiatto}, C. and {Tandy}, J. and {Teriaca}, L. and {Theobald}, C. and {Tychon}, I. and {van Driel-Gesztelyi}, L. and {Verbeeck}, C. and {Vial}, J. -C. and {Werner}, S. and {West}, M.~J. and {Westwood}, D. and {Wiegelmann}, T. and {Willis}, G. and {Winter}, B. and {Zerr}, A. and {Zhang}, X. and {Zhukov}, A.~N.},
        title = "{The Solar Orbiter EUI instrument: The Extreme Ultraviolet Imager}",
      journal = {\aap},
         year = 2020,
        month = oct,
       volume = {642},
          eid = {A8},
        pages = {A8},
          doi = {10.1051/0004-6361/201936663},
       adsurl = {https://ui.adsabs.harvard.edu/abs/2020A&A...642A...8R}
}

@ARTICLE{Bak_2013,
       author = {{B{\k{a}}k-St{\c{e}}{\'s}licka}, Urszula and {Gibson}, Sarah E. and {Fan}, Yuhong and {Bethge}, Christian and {Forland}, Blake and {Rachmeler}, Laurel A.},
        title = "{The Magnetic Structure of Solar Prominence Cavities: New Observational Signature Revealed by Coronal Magnetometry}",
      journal = {\apjl},
         year = 2013,
        month = jun,
       volume = {770},
       number = {2},
          eid = {L28},
        pages = {L28},
          doi = {10.1088/2041-8205/770/2/L28},
archivePrefix = {arXiv},
       eprint = {1304.7388},
 primaryClass = {astro-ph.SR},
       adsurl = {https://ui.adsabs.harvard.edu/abs/2013ApJ...770L..28B}
}

@ARTICLE{Ruminiska_2022,
       author = {{Rumi{\'n}ska}, Agnieszka and {B{\k{a}}k-St{\c{e}}{\'s}licka}, Urszula and {Gibson}, Sarah E. and {Fan}, Yuhong},
        title = "{Coronal Cavities in CoMP Observations}",
      journal = {\apj},
         year = 2022,
        month = feb,
       volume = {926},
       number = {2},
          eid = {146},
        pages = {146},
          doi = {10.3847/1538-4357/ac469c},
       adsurl = {https://ui.adsabs.harvard.edu/abs/2022ApJ...926..146R}
}

@ARTICLE{Gibson_streamer_2017,
       author = {{Gibson}, Sarah E. and {Dalmasse}, Kevin and {Rachmeler}, Laurel A. and {De Rosa}, Marc L. and {Tomczyk}, Steven and {de Toma}, Giuliana and {Burkepile}, Joan and {Galloy}, Michael},
        title = "{Magnetic Nulls and Super-radial Expansion in the Solar Corona}",
      journal = {\apjl},
         year = 2017,
        month = may,
       volume = {840},
       number = {2},
          eid = {L13},
        pages = {L13},
          doi = {10.3847/2041-8213/aa6fac},
archivePrefix = {arXiv},
       eprint = {1704.07470},
 primaryClass = {astro-ph.SR},
       adsurl = {https://ui.adsabs.harvard.edu/abs/2017ApJ...840L..13G}
}

@ARTICLE{Gilbert_2007,
       author = {{Gilbert}, Holly R. and {Alexander}, David and {Liu}, Rui},
        title = "{Filament Kinking and Its Implications for Eruption and Re-formation}",
      journal = {\solphys},
         year = 2007,
        month = oct,
       volume = {245},
       number = {2},
        pages = {287-309},
          doi = {10.1007/s11207-007-9045-z},
       adsurl = {https://ui.adsabs.harvard.edu/abs/2007SoPh..245..287G}
}

@ARTICLE{susino_2018,
       author = {{Susino}, R. and {Bemporad}, A. and {Jej{\v{c}}i{\v{c}}}, S. and {Heinzel}, P.},
        title = "{Hot prominence detected in the core of a coronal mass ejection. III. Plasma filling factor from UVCS Lyman-{\ensuremath{\alpha}} and Lyman-{\ensuremath{\beta}} observations}",
      journal = {\aap},
         year = 2018,
        month = sep,
       volume = {617},
          eid = {A21},
        pages = {A21},
          doi = {10.1051/0004-6361/201832792},
archivePrefix = {arXiv},
       eprint = {1805.12465},
 primaryClass = {astro-ph.SR},
       adsurl = {https://ui.adsabs.harvard.edu/abs/2018A&A...617A..21S}
}

@ARTICLE{Choudhary_2003,
       author = {{Choudhary}, Debi Prasad and {Moore}, Ronald L.},
        title = "{Filament eruption without coronal mass ejection}",
      journal = {\grl},
         year = 2003,
        month = nov,
       volume = {30},
       number = {21},
          eid = {2107},
        pages = {2107},
          doi = {10.1029/2003GL018332},
       adsurl = {https://ui.adsabs.harvard.edu/abs/2003GeoRL..30.2107C}
}

@ARTICLE{Seki_2021,
       author = {{Seki}, Daikichi and {Otsuji}, Kenichi and {Ishii}, Takako T. and {Asai}, Ayumi and {Ichimoto}, Kiyoshi},
        title = "{Relationship between three-dimensional velocity of filament eruptions and CME association}",
      journal = {Earth, Planets and Space},
         year = 2021,
        month = dec,
       volume = {73},
       number = {1},
          eid = {58},
        pages = {58},
          doi = {10.1186/s40623-021-01378-4},
archivePrefix = {arXiv},
       eprint = {2102.04578},
 primaryClass = {astro-ph.SR},
       adsurl = {https://ui.adsabs.harvard.edu/abs/2021EP&S...73...58S}
}

@ARTICLE{Schmieder_2013,
       author = {{Schmieder}, B. and {D{\'e}moulin}, P. and {Aulanier}, G.},
        title = "{Solar filament eruptions and their physical role in triggering coronal mass ejections}",
      journal = {Advances in Space Research},
         year = 2013,
        month = jun,
       volume = {51},
       number = {11},
        pages = {1967-1980},
          doi = {10.1016/j.asr.2012.12.026},
archivePrefix = {arXiv},
       eprint = {1212.4014},
 primaryClass = {astro-ph.SR},
       adsurl = {https://ui.adsabs.harvard.edu/abs/2013AdSpR..51.1967S}
}

@ARTICLE{Yan_2011,
       author = {{Yan}, X. -L. and {Qu}, Z. -Q. and {Kong}, D. -F.},
        title = "{Relationship between eruptions of active-region filaments and associated flares and coronal mass ejections}",
      journal = {\mnras},
         year = 2011,
        month = jul,
       volume = {414},
       number = {4},
        pages = {2803-2811},
          doi = {10.1111/j.1365-2966.2011.18336.x},
archivePrefix = {arXiv},
       eprint = {1101.3625},
 primaryClass = {astro-ph.SR},
       adsurl = {https://ui.adsabs.harvard.edu/abs/2011MNRAS.414.2803Y}
}

@ARTICLE{SMM1980,
       author = {{MacQueen}, R.~M. and {Csoeke-Poeckh}, A. and {Hildner}, E. and {House}, L. and {Reynolds}, R. and {Stanger}, A. and {Tepoel}, H. and {Wagner}, W.},
        title = "{The High Altitude Observatory coronagraph/polarimeter on the Solar Maximum Mission.}",
      journal = {\solphys},
         year = 1980,
        month = feb,
       volume = {65},
       number = {1},
        pages = {91-107},
          doi = {10.1007/BF00151386},
       adsurl = {https://ui.adsabs.harvard.edu/abs/1980SoPh...65...91M}
}

@ARTICLE{Song_2017,
       author = {{Song}, H.~Q. and {Cheng}, X. and {Chen}, Y. and {Zhang}, J. and {Wang}, B. and {Li}, L.~P. and {Li}, B. and {Hu}, Q. and {Li}, G.},
        title = "{The Three-part Structure of a Filament-unrelated Solar Coronal Mass Ejection}",
      journal = {\apj},
         year = 2017,
        month = oct,
       volume = {848},
       number = {1},
          eid = {21},
        pages = {21},
          doi = {10.3847/1538-4357/aa8d1a},
       adsurl = {https://ui.adsabs.harvard.edu/abs/2017ApJ...848...21S}
}

@ARTICLE{Filippov_2020,
       author = {{Filippov}, B.},
        title = "{Failed prominence eruptions near 24 cycle maximum}",
      journal = {\mnras},
         year = 2020,
        month = may,
       volume = {494},
       number = {2},
        pages = {2166-2177},
          doi = {10.1093/mnras/staa896},
archivePrefix = {arXiv},
       eprint = {2003.12988},
 primaryClass = {astro-ph.SR},
       adsurl = {https://ui.adsabs.harvard.edu/abs/2020MNRAS.494.2166F}
}

@ARTICLE{Song_2025,
       author = {{Song}, Hongqiang and {Wang}, Rui and {Li}, Leping and {Wang}, Bing and {Chen}, Yao},
        title = "{On the Nature of the Bright Front of Solar Coronal Mass Ejections}",
      journal = {\apj},
         year = 2025,
        month = aug,
       volume = {988},
       number = {2},
          eid = {270},
        pages = {270},
          doi = {10.3847/1538-4357/adec88},
       adsurl = {https://ui.adsabs.harvard.edu/abs/2025ApJ...988..270S}
}
\bibliographystyle{aasjournalv7}



\end{document}